\RequirePackage{etoolbox}
\csdef{input@path}{{style/}{graphics/}}
\documentclass[aap,MSNbibl,seceqn,dvips]{arximspdf}
\usepackage{mathbh}
\usepackage{graphicx}

\doi{10.1214/14-AAP1025} 
\volume{25}
\issue{3}
\pubyear{2015}
\firstpage{1349}
\lastpage{1382}

\makeatletter
\def\widetilde{\tilde}

\newcommand{\rrvert}{\vert}
\newcommand{\llvert}{\vert}
\def\cal{\mathcal}
\newcommand{\eqref}[1]{(\ref{#1})}
\newcommand{\T}{\textsf{\textit{x}}}
\newcommand{\Or}{\mathcal{O}}
\newcommand{\Ai}{\mathrm{Ai}}
\newcommand{\const}{\operatorname{const}}
\newcommand{\Pb}{\mathbb{P}}
\newcommand{\Id}{\mathbh{1}}
\newcommand{\e}{\varepsilon}
\newcommand{\I}{\mathrm{i}}
\newcommand{\D}{\mathrm{d}}
\newcommand{\C}{\mathbb{C}}
\newcommand{\R}{\mathbb{R}}
\newcommand{\Z}{\mathbb{Z}}
\renewcommand{\Re}{\operatorname{Re}}
\renewcommand{\Im}{\operatorname{Im}}
\newcommand{\sgn}{\operatorname{sgn}}

\newtheorem{prop}{Proposition}[section]
\newtheorem{thmm}[prop]{Theorem}
\newtheorem{lem}[prop]{Lemma}
\newproclaim{defin}[prop]{Definition}
\newtheorem{cor}[prop]{Corollary}

\newproclaim{rem}[prop]{Remark}
\newproclaim{example}[prop]{Example}
\makeatother

\begin{document}
\begin{frontmatter}

\title{Scaling limit for Brownian motions with one-sided~collisions}
\runtitle{Scaling limit for Brownian motions}

\begin{aug}
\author[A]{\fnms{Patrik L.}~\snm{Ferrari}\thanksref{T1}\ead[label=e1]{ferrari@uni-bonn.de}},
\author[B]{\fnms{Herbert}~\snm{Spohn}\ead[label=e2]{spohn@ma.tum.de}}
\and
\author[B]{\fnms{Thomas} \snm{Weiss}\corref{}\thanksref{T2}\ead[label=e3]{tweiss@ma.tum.de}\ead[label=u1,url]{http://www.foo.com}}
\thankstext{T1}{Supported by the German Research Foundation via the
SFB 1060--B04 project.}
\thankstext{T2}{Supported by the ENB graduate program, ``TopMath.''}
\runauthor{P.~L. Ferrari, H. Spohn and T. Weiss}
\affiliation{Bonn University, TU M\"unchen and TU M\"unchen}
\address[A]{P.~L. Ferrari\\
Institute for Applied Mathematics\\
Bonn University\\
Endenicher Allee 60\\
53115 Bonn\\
Germany\\
\printead{e1}}
\address[B]{H. Spohn\\
T. Weiss\\
Zentrum Mathematik\\
TU M\"unchen\\
Boltzmannstrasse 3\\
D-85747 Garching\\
Germany\\
\printead{e2}\\
\phantom{E-mail:\ }\printead*{e3}}
\end{aug}

\received{\smonth{7} \syear{2013}}
\revised{\smonth{11} \syear{2013}}

%
\begin{abstract}
We consider Brownian motions with one-sided collisions, meaning that
each particle is reflected at its right neighbour. For a finite number
of particles a Sch\"utz-type formula is derived for the transition
probability. We investigate an infinite system with periodic initial
configuration, that is, particles are located at the integer lattice at
time zero. The joint distribution of the positions of a finite subset
of particles is expressed as a Fredholm determinant with a kernel
defining a signed determinantal point process. In the appropriate large
time scaling limit, the fluctuations in the particle positions are
described by the Airy$_1$ process.
\end{abstract}

%
%
\end{frontmatter}
%
\section{Introduction}\label{SectIntro}
A widely studied model of interacting Brownian motions is governed by
the coupled stochastic differential equations
%
\begin{equation}
\label{1} \mathrm{d}x_j = \bigl(V'(x_{j+1} -
x_j) - V'(x_{j} - x_{j-1})\bigr)\,\mathrm{d}t + \sqrt{2} \,\mathrm{d}B_j(t),
\end{equation}
$j = 1,\ldots,N$, written here for the case where particles diffuse in
one dimension. Hence $x_j(t) \in\mathbb{R}$ and $\{B_j(t),j =
1,\ldots,N\}$ is a collection of $N$ independent standard Brownian motions. The
boundary terms $V'(x_{N+1} - x_N)$ and $V'(x_{1} - x_0)$ are to be set
equal to 0. The solutions to (\ref{1})
define a reversible diffusion process in $\mathbb{R}^N$ with respect to
the stationary measure
%
\begin{equation}
\label{2} \exp \Biggl(-\sum_{j=1}^{N-1}
V(x_{j+1} - x_j) \Biggr) \prod
_{j=1}^N \,\mathrm{d}x_j.
\end{equation}
Particle $j$ interacts with both, right and left, neighboring particles
with labels $j+1$ and $j-1$.

In our contribution we will study the case where the interaction is
only with the right neighbor. Hence, including an adjustment of the
noise strength,
%
\begin{equation}
\label{3} \mathrm{d}x_j = V'(x_{j+1} -
x_j) \,\mathrm{d}t + \mathrm{d}B_j(t),
\end{equation}
$j = 1,\ldots,N$. Somewhat unexpectedly, the measure (\ref{2}) is still
stationary. Of course, now the diffusion process is no longer
reversible. As to be discussed this modification will change
dramatically the large scale properties of the dynamics.

A special case is the exponential potential $e^{-\beta x}$, $\beta>
0$, which is related to quantum Toda chains,
Gelfand--Tsetlin patterns and other structures from quantum integrable
systems~\cite{BC11,OCon09}. Our focus is the hard collision limit,
\mbox{$\beta\to\infty$}. Then the positions will be ordered as $x_N
\leq\cdots\leq x_1$. Hence the diffusion process $x(t)$
has the Weyl chamber $\mathbb{W}_N = \{x | x_N \leq\cdots\leq x_1\}
$ as state space. Away from $\partial\mathbb{W}_N $,
$x(t)$ is simply $N$-dimensional Brownian motion. The interactions are
point-like and particle $j+1$ is reflected from particle $j$. These are
the one-sided collisions of the title. As a rare circumstance, for
every $N$ this diffusion process possesses an
explicit Sch\"{u}tz-type formula for its transition probability~\cite
{Sch97,War07}. For the particular initial condition $x(0) = 0$,
it follows from the Sch\"{u}tz-type formula that $x_N(t)$ has the same
distribution as the largest eigenvalue
of a $N\times N$ GUE random matrix. Even stronger, the process $t \to
-x_N(t)$ has the same law as the top line
of $N$-particle Dyson's Brownian motion starting at $0$~\cite
{Bar01,TW94}. It then follows that
%
\begin{equation}
\label{4} \lim_{t \to\infty} \frac{1}{\sigma t^{1/3}} \bigl(
x_{\lfloor at
\rfloor
}(t) - \mu t \bigr) = \xi_{\mathrm{GUE}}.
\end{equation}
Here $\lfloor\cdot\rfloor$ denotes integer part. The coefficients
$\sigma,\mu$ depend on $a>0$, and
$\xi_{\mathrm{GUE}}$ is a Tracy--Widom distributed random variable. One
can also consider the particle label
$\lfloor at + rt^{2/3} \rfloor$. Then in (\ref{4}) one has a stochastic
process in $r$ and it converges to the $\mathrm{Airy}_2$ process~\cite
{FN08b}. Alternatively, one could consider the label $\lfloor at
\rfloor
$, but different times $t + rt^{2/3}$, resulting in the same limit
process~\cite{Wei11}. This can also be derived from the fixed time
result using the slow decorrelations along characteristics~\cite{Fer08,CFP10b}.

In our contribution we will investigate the equally spaced initial
condition $x_j(0) = -j$, $j\in\Z$. Our main result is that the limit
(\ref{4}) still holds provided $\xi_{\mathrm{GUE}}$ is replaced by
$\xi
_{\mathrm{GOE}}$, that is, the Tracy--Widom distribution for a Gaussian
Orthogonal Ensemble. Also the $\mathrm{Airy}_2$ process will have to be
replaced by the $\mathrm{Airy}_1$ process; see Theorem~\ref
{thmAsympFixedTime}.

The limit (\ref{4}) can also be studied for the reversible process
governed by (\ref{1}). In this case other methods are available, listed
under the heading of nonequilibrium hydrodynamic fluctuation
theory~\cite{CY92}, which work for a large class of potentials~$V$.
Then $t^{1/3}$ would have to be replaced by $t^{1/4}$ and $\xi
_{\mathrm
{GUE}}$ by a Gaussian random variable. In this case the hard collision
limit corresponds to independent Brownian particles with the order of
particle labels maintained. The $t^{1/4}$ behavior is a famous result
by Harris \cite{Har65}. For nonreversible diffusion processes, as in
(\ref{3}), one is still limited to a very special choice of $V$. But it
is expected that the result holds in greater generality for a large
class of potentials.

For the one-sided collision limit, the solution to (\ref{3}) can be
represented as a last passage problem, which has the same structure as
directed polymers at zero temperature~\cite{OCon09,OCY01}. Also, (\ref
{3}) can be viewed as a particular discretization of the KPZ
equation~\cite{KPZ86}. While these links help us to come up with
convincing conjectures, our proof uses disjoint methods by relying on
the special structure of the transition probability. The same structure
is familiar from the TASEP with periodic initial conditions as has been
investigated in~\cite{BFP06,BFPS06,Sas05}. Some constructions developed
there carry over directly to our case. But novel steps are needed, like
the bi-orthogonalization in our set-up. Also the Lambert function
apparently has not made its appearance before.

\section{Model and main results}\label{SectMainResults}
One way to define a Brownian motion, $\T(t)$, starting from $\T(0)\in
\R
$ and being reflected at some continuous function $f(t)$ with $f(0)<\T
(0)$ is via the Skorokhod representation~\cite{Sko61,AO76}
%
\begin{eqnarray}
\T(t)&=&\T(0)+B(t)-\min \Bigl\{0,\inf
_{0\leq s \leq t}\bigl(\T (0)+B(s)-f(s)\bigr) \Bigr\}
\nonumber
\\[-8pt]
\\[-8pt]
\nonumber
&=& \max \Bigl\{\T(0)+B(t),\sup_{0\leq s \leq t}\bigl(f(s)+B(t)-B(s)
\bigr) \Bigr\},
\end{eqnarray}
where $B$ is a standard Brownian motion starting at $0$.

Let $B_k$, $k\in\Z$, be independent standard Brownian motions starting
at $0$, and define the random variables
%
\begin{equation}
\label{eq2.2} Y_{k,m}(t)=\sup_{0\leq s_{k+1}\leq\cdots\leq s_m\leq t}\sum
_{i=k}^m \bigl(B_i(s_{i+1})-B_i(s_i)
\bigr)
\end{equation}
with the convention $s_{k}=0$ and $s_{m+1}=t$.

Then, iterating the Skorokhod representation, we can define $N$
Brownian motions, $\T_1,\ldots,\T_N$, starting at positions $\T
_1(0)\geq\T_2(0)\geq\cdots\geq\T_N(0)$, such that the Brownian
motion $\T_k$ is reflected at the trajectory of Brownian motion $\T
_{k-1}$ according to
%
\begin{equation}
\label{eq3} \T_m(t)=-\max_{1\leq k\leq m}\bigl
\{Y_{k,m}(t)-\T_k(0)\bigr\}, \qquad 1\leq m \leq N.
\end{equation}
This is a Brownian motion in the $N$-dimensional Weyl chamber with $\pi
/4$ oblique reflections~\cite{WW84,KPS12,HW87}. Equivalently we
visualize the dynamics as $N$ Brownian particles in $\mathbb{R}$
interacting through one-sided collisions.
The process $\{-\T_1(t),\ldots,-\T_N(t)\}$ can be also interpreted as
the zero-temperature O'Connell--Yor semi-directed polymer model~\cite
{OCY01,OCon09} modified by assigning the extra weights $-\T_1(0),\T
_1(0)-\T_2(0),\ldots,\T_{N-1}(0)-\T_N(0)$ at time $0$.

An equivalent description is given by
%
\begin{eqnarray}
\T_1(t)&=&\T_1(0)+B_1(t),
\nonumber
\\[-8pt]
\\[-8pt]
\nonumber
\T_m(t)&=&\T_m(0)+B_m(t)-L_{\T_{m-1}-\T_m}(t),\qquad
m=2,\ldots,N,
\end{eqnarray}
where $L_{X-Y}(t)$ is twice the semimartingale local time at zero of
$X(t)-Y(t)$. This point of view is used in~\cite{War07}, where Warren
obtained a formula for the transition density of the system with $N$
Brownian motions (Proposition~8 of~\cite{War07}, reported as
Proposition~\ref{PropWarren} below). His result will be the starting
point for our analysis.

In this contribution we consider the case of infinitely many Brownian
particles starting from fixed, equally spaced positions, which w.l.o.g.
we set it to be $1$. This system is obtained as a limit of the
following system of finitely many Brownian particles.
Let us denote by
%
\begin{equation}
\label{eq2.5} x_m^{(M)}(t)=-\max_{k\in[-M+1,m]}
\bigl\{Y_{k,m}(t)+k\bigr\},
\end{equation}
for $m\in[-M+1,M]$. This defines the system of $2M$ reflected Brownian
particles starting at time zero from $x_m^{(M)}(0)=-m$. The $M\to
\infty
$ limit of this process is well defined in the sense that the
trajectories of finitely many of them converge in uniform norm over any
finite time interval; see Section~\ref{SectInfiniteParticles} for the proof.

\begin{prop}\label{propMconv}
Let us define
%
\begin{equation}
x_m(t)=-\max_{k\leq m}\bigl\{Y_{k,m}(t)+k
\bigr\}.
\end{equation}
Then for any $T>0$,
%
\begin{equation}
\lim_{M\to\infty}\sup_{t\in[0,T]}\bigl|x_m^{(M)}(t)-x_m(t)\bigr|=0\qquad
\mbox{a.s.}
\end{equation}
as well as
%
\begin{equation}
\sup_{t\in[0,T]}\bigl|x_m(t)\bigr|<\infty \qquad\mbox{a.s.}
\end{equation}
\end{prop}

As a first main result, we provide an expression for the joint
distribution at fixed time $t$.

\begin{figure}[b]

\includegraphics{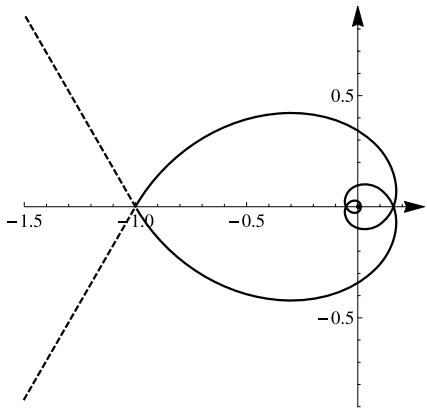}

\caption{(Dashed line) a possible choice for the contour
$\Gamma_-$ and (solid line) its image by~$\varphi$.}\label{FigContoursProp}
\end{figure}

\begin{prop}\label{propFlat}
Consider the initial condition with infinitely many Brownian motions,
indexed by $k\in\Z$, starting at positions $x_k(0)=-k$. Then, for any
finite subset $S$ of $\Z$, it holds
%
\begin{equation}
\label{eq33} \Pb \biggl(\bigcap_{k\in S} \bigl
\{x_k(t)\geq a_k\bigr\} \biggr)=\det\bigl(
\Id-P_a K_t^{\mathrm{flat}} P_a
\bigr)_{L^2(\R\times S)},
\end{equation}
where $P_a(x,k)=\Id_{(-\infty,a_k)}(x)$ and the kernel $K_t^{\mathrm{flat}}$
is given by
%
\begin{eqnarray}
\label{eqKtflat} %
K_t^{\mathrm{flat}}(x_1,n_1;x_2,n_2)&=&-
\frac
{(x_1-x_2)^{n_2-n_1-1}}{(n_2-n_1-1)!}\Id(x_1\geq x_2)\Id(n_2>n_1)
\nonumber
\\[-8pt]
\\[-8pt]
\nonumber
&&{}+\frac{1}{2\pi\I} \int_{\Gamma_-}\, \D z\frac{e^{t z^2/2} e^{-z
x_1}(-z)^{n_1}}{e^{t \varphi(z)^2/2} e^{-\varphi(z) x_2} (-\varphi
(z))^{n_2}}.
\end{eqnarray}
Here $\Gamma_-$ is any path going from $\infty e^{-\theta\I}$ to
$\infty e^{\theta\I}$ with $\theta\in[\pi/2,3\pi/4)$, crossing the
real axis to the left of $-1$, and such that the function
%
\begin{equation}
\varphi(z)=L_0\bigl(z e^{z}\bigr)
\end{equation}
is continuous and bounded. Here $L_0$ is the Lambert W function, that
is, the principal solution for $w$ in $z=w e^w$; see Figure~\ref
{FigContoursProp}.
\end{prop}

%

Interesting and quite unexpected is the appearance of the Lambert
function, defined as the multivalued inverse of the function $z\mapsto
ze^z$. It has a branch structure similar to the logarithm, but slightly
more complicated. The Lambert function is of use in many different
areas like combinatorics, exponential towers, delay-differential
equations~\cite{Che02} and several problems from physics~\cite
{BJ00,JK04,CJV00}. This function has been studied in detail; for
example, see~\cite{BPLPCS00,CJK97,CGHJ93}, with~\cite{CGHJK96} the
standard reference. However, the specific behavior needed for our
asymptotic analysis does not seem to be covered in the literature.

\textit{Equal time limit process.}
As second main result of our contribution we provide a characterization
of the law for the positions of the interacting Brownian motions in the
large time limit. Due to the asymmetric reflections, the particles have
an average velocity $-1$. For large time $t$ the KPZ scaling theory
suggests the positional fluctuations relative to the characteristic to
be of order $t^{1/3}$. Nontrivial correlations between particles occur
if the particle indices are of order $t^{2/3}$ apart from each other.
Therefore, to describe the Brownian particles close to the origin at
time $t$, we consider the scaling of the labels as
%
\begin{equation}
n(r,t)=\bigl\lfloor-t+2^{5/3}t^{2/3}r\bigr\rfloor,
\end{equation}
and we define the rescaled process as
%
\begin{equation}
\label{defXt} r\mapsto X_t(r)=-\frac{x_{n(r,t)}(t)+2^{5/3}t^{2/3}r}{(2t)^{1/3}}.
\end{equation}
The limit object is the Airy$_1$ process, which is defined as follows.

\begin{defin}
Let $B_0(x,y)=\Ai(x+y)$, with $\Ai$ the standard Airy function,
$\Delta
$ the one-dimensional Laplacian and the kernel $K_{\mathcal{A}_1}$
defined by
%
\begin{eqnarray}
\label{defairykernel} K_{\mathcal
{A}_1}(s_1,r_1;s_2,r_2)&=&-
\bigl(e^{(r_2-r_1)\Delta} \bigr) (s_1,s_2)\Id
(r_2>r_1)
\nonumber
\\[-8pt]
\\[-8pt]
\nonumber
&&{}+ \bigl(e^{-u_1\Delta}B_0e^{u_2\Delta}
\bigr) (s_1,s_2).
\end{eqnarray}
The \emph{Airy$_1$ process, $\mathcal{A}_1$}, is the process with
$m$-point joint distributions at $r_1<r_2<\cdots<r_m$ given by
the Fredholm determinant
%
\begin{equation}
\Pb \Biggl(\bigcap_{k=1}^m \bigl\{
\mathcal{A}_1(r_k)\leq s_k \bigr\} \Biggr)=
\det (\Id-\chi_s K_{\mathcal{A}_1}\chi_s
)_{L^2(\{
r_1,\ldots,r_m\}\times\R)},
\end{equation}
where $\chi_s(r_k,x)=\Id(x>s_k)$.
\end{defin}

Our second main result is the convergence of $X_t$ to the Airy$_1$ process.

%
\begin{thmm}\label{thmAsympFixedTime}
In the large time limit, $X_t$ converges to the Airy$_1$ process,
%
\begin{equation}
\lim_{t\to\infty}X_t(r) = \mathcal{A}_1(r),
\end{equation}
in the sense of finite-dimensional distributions.
\end{thmm}

Proposition~\ref{propFlat} is proved in Section~\ref{SectDetStructure}
and Theorem~\ref{thmAsympFixedTime}
in Section~\ref{SectAsympt}. Properties of the Lambert function are
collected in Appendix~\ref{AppendixLambert}.

\textit{Tagged particle limit process.}
The rescaled process at fixed time is not the only one in which the
Airy$_1$ process appears. It is also the case for the joint
distributions of the positions of a tagged Brownian motion at different
times. More precisely, consider the Brownian motion that started at the
origin at time $0$. Define its rescaled position by
%
\begin{equation}
\tau\mapsto X^{\mathrm{tagged}}_t(\tau):= -\frac{x_0(t+\tau
2^{5/3}t^{2/3})+ (t+\tau2^{5/3} t^{2/3})}{(2t)^{1/3}}.
\end{equation}
This rescaled process converges to the Airy$_1$ process.

\begin{thmm}\label{thmTagged}
In the large time limit,
%
\begin{equation}
\lim_{t\to\infty} X^{\mathrm{tagged}}_t(\tau) =
\mathcal{A}_1(\tau),
\end{equation}
in the sense of finite-dimensional distributions.
\end{thmm}

This theorem is proven in Section~\ref{SectTagged}. It is a special
case of the more general statement of Theorem~\ref{ThmAsymptSpaceTime}
in Section~\ref{SectTagged}. The result is based from the fixed time
result, Theorem~\ref{thmAsympFixedTime}, and a slow decorrelation
result, Proposition~\ref{PropSlowDecorrelation}. The latter says that
along special space--time directions the decorrelation happens over a
macroscopic time span.

\subsection*{Attractiveness and a more general class of initial data}
A stochastic particle system is called attractive, if for two distinct
initial configurations evolving under the same noise their order is
preserved. This property is shared by our model.

\begin{prop}\label{propBoundMod}
Let us consider two initial conditions, denoted by $\{a_m\}_{m \in
\mathbb{Z}}$, $\{b_m\}_{m\in\mathbb{Z}}$. Under the same
noise they evolve to $x_m^a(t)$ and $x_m^b(t)$. If there is $M>0$ such
that $|a_m-b_m|\leq M$ for all $m\in\mathbb{Z}$, then also
%
\begin{equation}
\bigl|x_m^a(t)-x_m^b(t)\bigr|\leq M\qquad
\forall m\in\mathbb{Z}, t>0.
\end{equation}
\end{prop}

The same property holds for the standard coupling of the TASEP, as
explained in Section~2.1 of~\cite{BFS07}.

As an immediate consequence, the limit result of Theorem~\ref{thmAsympFixedTime}
holds for
bounded modifications of the initial condition $x_m(0)=-m$, since an
error of size $M$ vanishes under the $t^{1/3}$ scaling. For example,
one could choose a unit cell of length 1
and take an arbitrary initial condition with the only restriction that
there are $\ell$ particles in each cell. Then the convergence to the
Airy$_1$ process holds.

\begin{pf*}{Proof of Proposition~\ref{propBoundMod}}
By definition,
%
\begin{eqnarray}
 x_m^a(t)&=&-\max
_{ k\leq m}\bigl\{Y_{k,m}(t)-a_k\bigr\},
\nonumber
\\[-8pt]
\\[-8pt]
\nonumber
x_m^b(t)&=&-\max_{ k\leq m}\bigl
\{Y_{k,m}(t)-b_k\bigr\}.
\end{eqnarray}
Since the inequality
%
\begin{equation}
Y_{k,m}(t)-a_k\leq Y_{k,m}(t)-b_k+M
\end{equation}
holds for each $k$, the maximum can be taken on each side, resulting in
%
\begin{eqnarray}
\max_{ k\leq m}\bigl
\{Y_{k,m}(t)-a_k\bigr\}&\leq&\max_{ k\leq m}
\bigl\{ Y_{k,m}(t)-b_k\bigr\} + M,
\nonumber
\\[-8pt]
\\[-8pt]
\nonumber
x_m^a(t)&\geq& x_m^b(t)-M.
\end{eqnarray}
Correspondingly, one has $x_m^b(t)\geq x_m^a(t)-M$.
\end{pf*}

\section{Limit to infinitely many Brownian particles}\label
{SectInfiniteParticles}
In this section we prove Proposition~\ref{propMconv}.
Given standard independent Brownian motions $B_{-M+1},\ldots,\break B_{M}$ we
define as in (\ref{eq2.2})--(\ref{eq2.5}),
%
\begin{equation}
Y_{k,m}(t)=\sup_{0\leq s_{k+1}\leq\cdots\leq s_m\leq t}\sum
_{i=k}^m \bigl(B_i(s_{i+1})-B_i(s_i)
\bigr)
\end{equation}
and
%
\begin{eqnarray}
x_m^{(M)}(t)&=&-\max
_{-M+1\leq k\leq m}\bigl\{Y_{k,m}(t)+k\bigr\},
\nonumber
\\[-8pt]
\\[-8pt]
\nonumber
x_m(t)&=&-\max_{ k\leq m}\bigl\{Y_{k,m}(t)+k
\bigr\}.
\end{eqnarray}

For the proof of Proposition~\ref{propMconv} we use following
concentration inequality result.

\begin{prop}[(Proposition~2.1 of \cite{ledDIL})]\label{concIn}
For each $T>0$ there exists a constant $C>0$ such that for all \mbox
{$k<m$}, $\delta>0$,
%
\begin{equation}
\Pb \biggl(\frac{Y_{k,m}(T)}{2\sqrt{(m-k+1)T}}\geq1+\delta \biggr)\leq Ce^{-(m-k+1)\delta^{3/2}/C}.
\end{equation}
\end{prop}

\begin{pf*}{Proof of Proposition~\ref{propMconv}}
Let
%
\begin{equation}
A_M:=\bigl\{Y_{-M,m}(T)-M\geq-M/2\bigr\}\cup\bigl
\{Y_{m,m}(T)+m\leq-M/2\bigr\}.
\end{equation}
We can deduce exponential decay of $\Pb(A_M)$ in $M$ from combining the
Gaussian tail of $Y_{m,m}$ with Proposition~\ref{concIn}, using
$\delta
=1$ and elementary inequalities. In particular $\sum_{M=1}^\infty\Pb
(A_M)<\infty$, so by Borel--Cantelli, $A_M$ occurs only finitely many
times almost surely. This means that a.s. there exists a $M'$, such
that for all $M\geq M'$,
%
\begin{eqnarray}
Y_{-M,m}(T)-M&<&-M/2 \quad\mbox{and}
\nonumber
\\[-8pt]
\\[-8pt]
\nonumber
Y_{m,m}(T)+m&>&-M/2.
\end{eqnarray}
Consequently, $Y_{m,m}(T)+m>Y_{-M,m}(T)-M$ for all $M\geq M'$ and therefore
%
\begin{equation}
x_m(T)=x_m^{(M')}(T)\qquad \mbox{a.s.}
\end{equation}

It remains to show uniformity over the time interval $[0,T]$. The above
argument implies that almost surely for any $t\in[0,T]$ there exists a
finite $M_t$ such that $x_m(t)=x_m^{(M_t)}(t)$.
Lemma~\ref{lemMaxPath} below implies that for any $t\in[0,T]$, it holds
$x_m(t)=x_m^{(M')}(t)$. This settles the convergence.

Finally we show that $\sup_{t\in[0,T]}|x_m^{(M')}(t)|<\infty$. This
follows from the bound
%
\begin{equation}
\bigl|Y_{k,m}(t)\bigr|\leq\sum_{i=k}^m
\Bigl(\sup_{0\leq s\leq t}B_i(s)-\inf_{0\leq s\leq t}B_i(s)
\Bigr)<\infty.
\end{equation}
\upqed\end{pf*}

\begin{lem}\label{lemMaxPath}
Consider $0\leq t_1\leq t_2$ and $m$, $M_{t_1}$, $M_{t_2}$ such that
%
\begin{equation}
\label{eq10a} x_m(t_i)=x_m^{(M_{t_i})}(t_i)\qquad
\mbox{for }i=1,2.
\end{equation}
Then
%
\begin{equation}
\label{eq11} x_m(t_1)=x_m^{(M_{t_2})}(t_1).
\end{equation}
\end{lem}

\begin{pf}
Define
%
\begin{equation}
S_M^m(a,b)= \bigl\{\mathbf{s}\in\R^{M+m+1}| a=
s_{-M+1}\leq\cdots \leq s_m\leq s_{m+1}=b \bigr\}.
\end{equation}
Notice that the definition of $Y_{k,m}$ contains a supremum of a
continuous function over the compact set $S_k^m(0,t)$, ensuring the
existence of a maximizing vector $\mathbf{s}$.

Another representation of $x_m^{(M)}(t)$ is
%
\begin{eqnarray}
x_m^{(M)}(t)&=&M-\sup
_{\mathbf{s}\in S_M^m(0,t)}\sum_{k=-M+1}^m
I_k,
\nonumber
\\[-8pt]
\\[-8pt]
\nonumber
I_k&=&B_k(s_{k+1})-B_k(s_k)+
\delta_{0,s_k}.
\end{eqnarray}
Notice that in \eqref{eq10a} we can replace $M_{t_i}$ by $M=\max\{
M_{t_1},M_{t_2}\}$. Condition \eqref{eq11} is equivalent to the
existence of a $\mathbf{s}^*\in S_M^m(0,t_1)$ such that $\sum_{k=-M+1}^m I_k$ is maximal and $s^*_{-M_{t_2}+1}=0$.

Let $\mathbf{s}^{(i)}\in S_M^m(0,t_i)$ be a maximizer of $\sum_{k=-M+1}^m I_k$. If $s^{(1)}_k\leq s^{(2)}_k$ for all $k$, then also
$s^{(1)}_{-M_{t_2}+1}\leq s^{(2)}_{-M_{t_2}+1}=0$, by \eqref{eq10a},
and the choice $\mathbf{s}^*=\mathbf{s}^{(1)}$ finishes the proof.

Otherwise let $k^*$ be the maximal $k$ such that $s^{(1)}_k>
s^{(2)}_k$. There exists $\tau$ with
%
\begin{equation}
s^{(2)}_{k^*}\leq s^{(1)}_{k^*}\leq\tau\leq
s^{(1)}_{k^*+1}\leq s^{(2)}_{k^*+1}.
\end{equation}
This allows the following decomposition:
%
\begin{eqnarray}
x_m^{(M_{t_i})}(t_i)&=&x_m^{(M)}(t_i)=M-
\sup_{\mathbf{s}\in
S_M^m(0,t_i)}\sum_{k=-M+1}^m
I_k
\nonumber
\\[-8pt]
\\[-8pt]
\nonumber
&=&M-\sup_{\mathbf{s}\in S_M^{k^*}(0,\tau)}\sum_{k=-M+1}^{k^*}
I_k-\sup_{\mathbf{s}\in S_{-k^*+1}^m(\tau,t_i)}\sum_{k=k^*}^m
I_k.
\end{eqnarray}
Now the supremum over $S_M^{k^*}(0,\tau)$ is attained by both vectors
$(s_{-M+1}^{(i)},\ldots,\break s_{k^*}^{(i)},\tau)$. Consequently, $\sum_{k=-M}^m I_k$ is maximized also by
%
\begin{equation}
\mathbf{s}^*=\bigl(s_{-M+1}^{(2)},\ldots,s_{k^*}^{(2)},s_{k^*+1}^{(1)},
\ldots,s_{m+1}^{(1)}\bigr),
\end{equation}
satisfying $s^*_{-M_{t_2}+1}=s^{(2)}_{-M_{t_2}+1}=0$.
\end{pf}

\section{Determinantal structure of joint distributions}\label
{SectDetStructure}
Let us denote by\break $\T_1(t)>\T_2(t)>\cdots>\T_N(t)$ the positions
of the $N$ Brownian motions as defined in Section~\ref
{SectMainResults}. Their joint distribution has a density, denoted by
$G(x,t|\T(0))$,
%
\begin{equation}
\Pb \Biggl(\bigcap_{k=1}^N
\bigl\{\T_k(t)\in\D x_k\bigr\}\Big | \T _1(0),
\ldots,\T _N(0) \Biggr) = G\bigl(x,t|\T(0)\bigr) \prod
_{k=1}^N \,\D x_k.
\end{equation}
Warren~\cite{War07} proves an explicit formula for $G$.

\begin{prop}[(Proposition~8 of~\cite{War07})]\label{PropWarren}
The joint density of the positions of the reflected Brownian motions at
time $t$ starting from positions $\T_k(0)$, $k=1,\ldots,N$, is given by
%
\begin{equation}
\label{eq2} G\bigl(x,t|\T(0)\bigr)=\det\bigl(F_{i-j}
\bigl(x_{N+1-i}-\T_{N+1-j}(0),t\bigr)\bigr)_{1\leq i,j\leq N}
\end{equation}
with
%
\begin{equation}
\label{eqFk} F_k(x,t)=\frac{1}{2\pi\I} \int_{\I\R+\delta}
\,\D z \frac{e^{t z^2/2}
e^{-z x}}{z^k}
\end{equation}
for any $\delta>0$.
\end{prop}

\begin{pf}
Note that $X_k^k(t)$ in~\cite{War07} corresponds to $-\T_{k}(t)$ in
this paper. Hence the spatial coordinates are reversed. In
Proposition~8 of~\cite{War07} it is shown that
%
\begin{equation}
G\bigl(x,t|\T(0)\bigr)=\det\bigl(P_t^{(i-j)}
\bigl(x_j-\T_i(0)\bigr)\bigr)_{1\leq i,j\leq N}
\end{equation}
with
%
\begin{eqnarray}
P_t^{(0)}(x)&=&
\frac{1}{\sqrt{2\pi t}}e^{-x^2/(2t)},\nonumber
\\
P_t^{(-n)}(x)&=&(-1)^n\frac{\D^n}{\D x^n}P_t^{(0)}(x),\qquad
n\geq1,
\\
P_t^{(n)}(x)&=&\int_x^\infty\,
\D y \frac
{(y-x)^{n-1}}{(n-1)!}P_t^{(0)}(y),\qquad n\geq1. \nonumber
\end{eqnarray}
Using the identity
%
\begin{equation}
\label{eq6} \frac{1}{\sqrt{2\pi t}}e^{-x^2/(2t)} = \frac{1}{2\pi\I} \int
_{\I
\R
+\delta}\, \D z \,e^{t z^2/2} e^{-z x}
\end{equation}
(that holds for any $\delta$), we have $P_t^{(0)}(x)=F_0(x,t)$. Also,
we immediately get
$P_t^{(-n)}(x)=F_{-n}(x,t)$ for $n\geq1$. Further, for $\delta>0$ we have
%
\begin{eqnarray}
P^{(n)}_t(x)&=&\frac{1}{2\pi\I}
\int_{\I\R+\delta} \,\D z \,e^{t
z^2/2} \int_x^\infty
\,\D y \frac{(y-x)^{n-1}}{(n-1)!} e^{-z y}
\nonumber
\\[-8pt]
\\[-8pt]
\nonumber
&=&\frac{1}{2\pi\I}\int_{\I\R+\delta} \,\D z \frac{e^{t z^2/2} e^{-z
x}}{z^n} =
F_n(x,t)
\end{eqnarray}
for all $n\geq1$. Thus $G(x,t|\T(0))=\det(F_{i-j}(x_j-\T
_i(0),t))_{1\leq i,j\leq N}$, and the change of indices $(i,j)\to
(N+1-j,N+1-i)$ gives us (\ref{eq2}).
\end{pf}

Equation (\ref{eq2}) appeared previously in~\cite{SW98} too. A joint
distribution of the same form as in Proposition~\ref{PropWarren} occurs
also in the study of the totally asymmetric simple exclusion process
(TASEP)~\cite{Sch97} (reported as Lemma~3.1 in~\cite{BFPS06}).
Following the approach of Borodin et al.~\cite{BFPS06} for TASEP, we
can show that the joint distributions of the positions of the Brownian
particles can be expressed as a Fredholm determinant for a given
correlation kernel.

Using as a starting point Proposition~\ref{PropWarren} we prove the
result for finitely many Brownian particles starting at $\{
-N,-N+1,\ldots,-1\}$.

\begin{prop}\label{propFiniteN}
Consider the initial condition \mbox{$\T_k(0)=-k$} for $k=1,\ldots, N$.
Then, for any subset $S$ of $\{1,2,\ldots,N\}$, it holds
%
\begin{equation}
\Pb \biggl(\bigcap_{k\in S} \bigl\{\T_k(t)
\geq a_k\bigr\} \biggr)=\det(\Id -P_a K_t
P_a)_{L^2(\R\times S)},
\end{equation}
where $P_a(x,k)=\Id_{(-\infty,a_k)}(x)$ and the kernel $K_t$ is given by
%
\begin{equation}
K_t(x_1,n_1;x_2,n_2)=-
\phi^{(n_1,n_2)}(x_1,x_2)+\sum
_{k=1}^{n_2} \Psi ^{n_1}_{n_1-k}(x_1)
\Phi^{n_2}_{n_2-k}(x_2).
\end{equation}
Here
%
\begin{eqnarray}
 \phi^{(n_1,n_2)}(x_1,x_2)&=&
\frac
{(x_1-x_2)^{n_2-n_1-1}}{(n_2-n_1-1)!}\Id (x_1\geq x_2)\Id(n_2>n_1),
\nonumber\\
\Psi^n_{n-k}(x) &=& \frac{(-1)^{n-k}}{2\pi\I} \int
_{\I\R-\delta} \,\D z \,e^{t z^2/2} e^{-z (x+k)}z^{n-k},
\\
\Phi^n_{n-\ell}(x)&=&\frac{(-1)^{n-\ell}}{2\pi\I}\oint_{\Gamma
_0} \,\D
w \frac{e^{w(x+\ell)}}{e^{t w^2/2} w^{n-\ell}}\frac{1+ w}{w}\nonumber
\end{eqnarray}
for $\delta>0$.
\end{prop}

\begin{pf}
The proof is very similar to the one in~\cite{BFPS06}, except that now
space is continuous. We report in Appendix~\ref{AppendixA} the relevant
results from~\cite{BFPS06}. The straightforward but key identity is
%
\begin{equation}
\label{eq10} F_{n+1}(x,t)=\int_x^\infty\,
\D y F_n(y,t).
\end{equation}
Let us denote by $x_1^k:=x_k$, $k=1,\ldots,N$. The $k$th row of the
determinant of (\ref{eq2}) is given by
%
\begin{equation}
\bigl[F_{k-1}\bigl(x_1^{N+1-k}-x_N(0),t
\bigr) \cdots F_{k-N}\bigl(x_1^{N+1-k}-x_1(0),t
\bigr) \bigr].
\end{equation}
Using repeatedly identity (\ref{eq10}) we can rewrite this row as the
$(k-1)$th fold integral
%
\begin{eqnarray}
&&\int_{x_1^{N+1-k}}^\infty\,\D x_2^{N+2-k}
\cdots
\nonumber
\\[-8pt]
\\[-8pt]
\nonumber
&&\qquad{}\times\int_{x_{k-1}^{N-1}}^\infty\,\D x_k^N
\bigl[F_{0}\bigl(x_k^N-x_N(0),t
\bigr) \cdots F_{-N+1}\bigl(x_k^N-x_1(0),t
\bigr) \bigr].
\end{eqnarray}
We do this replacement to each row $k\geq2$, and by multi-linearity of
the determinant, we get
%
\begin{eqnarray}
\label{eq13} &&G\bigl(x,t|x(0)\bigr)
\nonumber
\\[-8pt]
\\[-8pt]
\nonumber
&&\qquad = \int_{\cal D'} \det
\bigl[F_{-j+1}\bigl(x_{i}^N-x_{N+1-j}(0),t
\bigr) \bigr]_{1\leq i,j\leq N}\prod_{2\leq
k \leq n\leq N}\,\D
x_k^n,
\end{eqnarray}
where the set $\cal D'$ is given by
%
\begin{equation}
{\cal D'}=\bigl\{x_k^n\in\R, 2\leq k \leq
n \leq N | x_k^n\geq x_{k-1}^{n-1}
\bigr\}.
\end{equation}
Then, using the antisymmetry in the variables $x_1^N,\ldots,x_N^N$ of
the determinant in~(\ref{eq13}), we can reduce the integration over
$\cal D$ (see Appendix~\ref{AppendixA}, Lemma~\ref{AppLemma3.3})
defined by
%
\begin{equation}
{\cal D}=\bigl\{x_k^n\in\R, 2\leq k \leq n \leq N |
x_k^n>x_k^{n+1},
x_k^n\geq x_{k-1}^{n-1}\bigr\}.
\end{equation}

The next step is to encode the constraint of the integration over $\cal
D$ into a formula and then consider the measure over $\{x_k^n,1\leq k
\leq n \leq N\}$, which turns out to have determinantal correlations
functions. At this point the allowed configurations are such that
$x_k^n\leq x_{k+1}^n$. For a while, we still consider ordered
configurations at each level, that is, with $x_1^n\leq x_2^n\leq\cdots
\leq x_n^n$ for $1\leq n\leq N$. Let us set
%
\begin{equation}
\widetilde{\cal D}=\bigl\{x_k^n\in\R, 1\leq k \leq n
\leq N | x_k^n>x_k^{n+1},
x_k^n\geq x_{k-1}^{n-1}\bigr\}.
\end{equation}
Defining \mbox{$\phi(x,y)=\Id(x> y)$}, it is easy to verify that
%
\begin{eqnarray}
&&\prod_{n=1}^{N-1} \det \bigl[\phi
\bigl(x_i^n,x_j^{n+1}\bigr)
\bigr]_{1\leq
i,j\leq n+1}
\nonumber
\\[-8pt]
\\[-8pt]
\nonumber
&&\qquad= \cases{ %
 1,&\quad$\mbox{if }
\bigl\{x_k^n,1\leq k \leq n \leq N\bigr\}\in\widetilde{
\cal D},$
\vspace*{2pt}\cr
0,&\quad $\mbox{otherwise},$}
\end{eqnarray}
where $x_{n+1}^n$ are ``virtual'' variables and $\phi(x_{n+1}^n,x):=1$.
We also set
%
\begin{equation}
\Psi^{N}_{N-k}(x):=(-1)^{N-k} F_{-N+k}
\bigl(x-x_k(0),t\bigr),
\end{equation}
for $k\in\{1,\ldots,N\}$. Then, (\ref{eq13}) can be obtained as a
marginal of the measure
%
\begin{equation}
\label{eq17} \frac{1}{Z_N} \prod_{n=1}^{N-1}
\det \bigl[\phi \bigl(x_i^n,x_j^{n+1}
\bigr) \bigr]_{1\leq i,j\leq n+1} \det \bigl[\Psi^N_{N-i}
\bigl(x_j^N\bigr) \bigr]_{1\leq
i,j\leq N}
\end{equation}
for some constant $Z_N$. Notice that the measure (\ref{eq17}) is
symmetric in the $x_k^n$'s since by permuting two of them (at the same
level $n$) one gets twice a factor $-1$. Thus, we relax the constraint
of ordered configurations at each level. The only effect is to modify
the normalization constant $Z_N$.

It is known from Lemma~3.4 of~\cite{BFPS06} (see Appendix~\ref
{AppendixA}, Lemma~\ref{AppLemma2}) that a (signed) measure of the form
(\ref{eq17}) has determinantal correlation functions, and the
correlation kernel is given as follows.
Let us set
%
\begin{equation}
\phi^{(n_1,n_2)}(x,y)= \cases{ %
\phi^{(*(n_2-n_1))}(x,y),&\quad $\mbox{if }n_1<n_2,$
\vspace*{2pt}\cr
0,&\quad $\mbox{if }n_1\geq n_2,$}
\end{equation}
and
%
\begin{equation}
\Psi^n_{n-k}(x):=\bigl(\phi^{(n,N)}*
\Psi^N_{N-k}\bigr) (x), \qquad\mbox{for }1\leq k\leq N.
\end{equation}
Assume that we have found families $\{\Phi^n_0(x),\ldots,\Phi
^n_{n-1}(x)\}$ such that $\Phi^n_k(x)$ is a polynomial of degree $k$
and they satisfy the biorthogonal relation
%
\begin{equation}
\label{eq21} \int_{\R}\, \D x \Psi^n_{n-k}(x)
\Phi^n_{n-\ell}(x)=\delta_{k,\ell
},\qquad 1\leq k,\ell\leq n.
\end{equation}
Then measure (\ref{eq17}) has correlation kernel given by
%
\begin{equation}
\label{eq22} \qquad K_t(n_1,x_1;n_2,x_2)=-
\phi^{(n_1,n_2)}(x_1,x_2)+\sum
_{k=1}^{n_2}\Psi ^{n_1}_{n_1-k}(x_1)
\Phi^{n_2}_{n_2-k}(x_2).
\end{equation}

Notice that in (\ref{eq13}) only $F_k$ with $k\leq0$ arises. In this
case, compare with (\ref{eqFk}), every sign of $\delta$ can be used, so
that by defining $\Psi^{N}_{N-k}$ above we decide to use the
integration path over $\I\R-\delta$, so that
%
\begin{equation}
\Psi^N_{N-k}(x)=\frac{(-1)^{N-k}}{2\pi\I} \int
_{\I\R-\delta} \,\D z \,e^{t
z^2/2} e^{-z (x-x_k(0))}z^{N-k},
\end{equation}
for any $\delta>0$. A simple computation gives [now we use $x_k(0)=- k$]
%
\begin{equation}
\Psi^n_{n-k}(x) = \frac{(-1)^{n-k}}{2\pi\I} \int
_{\I\R-\delta} \,\D z \,e^{t z^2/2} e^{-z (x+ k)}z^{n-k}.
\end{equation}
With a little bit of experience, it is not hard to find the
biorthogonal functions. They are given by
%
\begin{equation}
\label{eq25} \Phi^n_{n-\ell}(x)=\frac{(-1)^{n-\ell}}{2\pi\I}
\oint_{\Gamma_0} \,\mathrm{d}w \frac
{e^{w(x+\ell)}}{e^{t w^2/2} w^{n-\ell}}\frac{1+ w}{w}.
\end{equation}
Remark that the choice of the sign of $\delta$ in the definition of
$\Psi^N_{N-k}$ above is irrelevant for the biorthogonalization, since
there is no pole at $z=0$ for $k=1,\ldots,n$.
Indeed, (\ref{eq21}) can be written as
%
\begin{equation}
\label{eq26} \int_{\R_-} \,\D x \Psi^{n}_{n-k}(x)
\Phi^n_{n-\ell}(x) + \int_{\R_+} \,\D x
\Psi^{n}_{n-k}(x) \Phi^n_{n-\ell}(x).
\end{equation}
For the first term, we choose $\delta>0$ and the path $\Gamma_0$ for
$w$ satisfying $\Re(z-w)<0$. Then, we can take the integral over
$x$ inside, and we obtain
%
\begin{eqnarray}
&&\int_{\R_-} \,\D x \Psi^{n}_{n-k}(x)
\Phi^n_{n-\ell}(x)
\nonumber
\\[-8pt]
\\[-8pt]
\nonumber
&&\qquad=-\frac
{(-1)^{k-\ell}}{(2\pi\I)^2}\int_{\I\R-\delta}
\,\D z \oint_{\Gamma
_0}\,\D w \frac{e^{t z^2/2} z^{n-k}e^{- z k}}{e^{t w^2/2}w^{n-\ell}e^{- w
\ell
}}\frac{1+ w}{w(z-w)}.
\end{eqnarray}
For the second term, we choose $\delta<0$ and the path $\Gamma_0$ for
$w$ satisfying \mbox{$\Re(z-w)>0$}. Then we can take the integral over
$x$ inside, and we obtain the same expression up to a minus sign. The
net result of (\ref{eq26}) is a residue at $z=w$, which is given by
%
\begin{equation}\qquad
\frac{(-1)^{k-\ell}}{2\pi\I}\oint_{\Gamma_0}\,\D w \frac{1+ w}{w} \bigl(w
e^{ w}\bigr)^{\ell-k}=\frac{(-1)^{k-\ell}}{2\pi\I}\oint_{\Gamma_0}\,\D W
W^{\ell-k-1}=\delta_{k,\ell},
\end{equation}
where we made the change of variables $W=w e^{ w}$.
Finally, a simple computation gives
%
\begin{equation}
\label{eq30} \phi^{(n_1,n_2)}(x_1,x_2)=
\frac
{(x_1-x_2)^{n_2-n_1-1}}{(n_2-n_1-1)!}\Id (x_1\geq x_2)\Id(n_2>n_1),
\end{equation}
which has also the integral representations
%
\begin{eqnarray}
\label{eq31} %
\phi^{(n_1,n_2)}(x_1,x_2)
&= &\frac{1}{2\pi\I} \int_{\I\R-\delta
}\,\D z \frac{e^{-z(x_1-x_2)}}{(-z)^{n_2-n_1}}
\nonumber
\\[-8pt]
\\[-8pt]
\nonumber
&=&\frac{1}{2\pi\I} \oint_{\Gamma_0}\,\D w \frac
{e^{w(x_1-x_2)}}{w^{n_2-n_1}}
\Id(x_1\geq x_2)\Id(n_2>n_1).
\end{eqnarray}
\upqed\end{pf}

\begin{rem}\label{remarkFiniteKernel}
$\Phi^{n_2}_{n_2-k}(x)=0$ for $k>n_2$ since the pole at $w=0$ in (\ref
{eq25}) vanishes. Therefore we can extend the sum over $k$ to $\infty$.
If we choose the integration paths such that $|w e^{ w}|<|z e^{ z}|$,
then we can take the sum into the integrals and perform the (geometric)
sum explicitly, with the result
%
\begin{eqnarray}
\label{eq32}&& \sum_{k=1}^{n_2}
\Psi^{n_1}_{n_1-k}(x_1)\Phi^{n_2}_{n_2-k}(x_2)
\nonumber
\\[-8pt]
\\[-8pt]
\nonumber
&&\qquad=\frac{1}{(2\pi\I)^2} \int_{\I\R-\delta} \,\D z\oint _{\Gamma
_0} \,\D
w \frac{e^{t z^2/2} e^{-z x_1}(-z)^{n_1}}{e^{t
w^2/2} e^{-w x_2} (-w)^{n_2}}\frac{(1+ w) e^{ w}}{z e^{ z}-w e^{ w}}.
\end{eqnarray}
A possible choice of the paths such that $|w e^{ w}|<|z e^{ z}|$ is
satisfied is the following: $w=e^{\I\theta}/4$ with $\theta\in[-\pi,\pi
)$ and $z=-1+\I y$ with $y\in\R$.
\end{rem}

\begin{rem}
It is possible to reformulate $K_t^{\mathrm{flat}}$ in a slightly different
way. By doing the change of variables $w=\varphi(z)$, we get $\frac
{\D
z}{\D w}= \frac{(1+ w) e^{ w}}{(1+ z) e^{ z}}$. Let us denote by
$z_k(w)$, $k\in\Z$, the solutions of
%
\begin{equation}
z e^{ z}=w e^{ w}
\end{equation}
with the trivial one indexed by $z_0(w)=w$. Then
%
\begin{eqnarray}
\label{eq38} %
&&K_t^{\mathrm{flat}}(x_1,n_1;x_2,n_2)\nonumber\\
&&\qquad=-
\frac
{(x_1-x_2)^{n_2-n_1-1}}{(n_2-n_1-1)!}\Id(x_1\geq x_2)\Id(n_2>n_1)
\nonumber
\\[-8pt]
\\[-8pt]
\nonumber
&&\qquad\quad{}+\sum_{k\in\Z\setminus\{0\}}\frac{1}{2\pi\I} \oint_{\Gamma_0}
\,\D w\frac{e^{t z_k(w)^2/2} e^{-z_k(w) x_1}(-z_k(w))^{n_1}}{e^{t w^2/2}
e^{-w x_2} (-w)^{n_2}}\\
&&\hspace*{42pt}\qquad\quad{}\times \frac{(1+ w)e^{ w}}{(1+ z_k(w))e^{ z_k(w)}}. \nonumber
\end{eqnarray}
\end{rem}

\begin{rem}
The form of kernel (\ref{eq38}) can be also derived by considering the
low density totally asymmetric simple exclusion process (TASEP). One
considers the initial condition of particles starting every $d$
position, that is, with density $\rho=1/d$. The kernel for this system
is given in~\cite{BFP06}, Theorem~2.1, where one should, however,
replace $(1+p u_i(v))^t/(1+pv)^t$ by $e^{u_i(v) t}/e^{v t}$ since
in~\cite{BFP06} a discrete time model was considered. Then taking the
$d\to\infty$ limit, with space and time rescaled diffusively, one
recovers (\ref{eq38}).
\end{rem}

\begin{pf*}{Proof of Proposition~\ref{propFlat}}
The idea is to consider the finite system, replace $x_i$ by $x_i- M$
and $n_i$ by $n_i+M$, and then take the $M\to\infty$ limit. The part of
the kernel which should survive the limit is the $M$-independent part.
The reformulation of Remark~\ref{remarkFiniteKernel} can be used, but
it is not the best for our purpose. Instead, notice that the path used
in $\Psi^n_{n-k}$ does not have necessarily be vertical. We can take
any path passing to the left of $0$ and such that it asymptotically
have an angle between in $(\pi/4,3\pi/4)$. In that case, the quadratic
term in $z$ is strong enough to ensure convergence of the integral.
Thus, we choose the path for
%
\begin{equation}
\label{eq36a} z\in\Gamma_-:=\bigl\{-2+e^{2\pi\I/3 \sgn(y)}|y|,y\in\R\bigr\}
\end{equation}
and
%
\begin{equation}
\label{eq36} w\in\Gamma_0:=\bigl\{e^{\I\theta},\theta\in[-\pi,
\pi)\bigr\};
\end{equation}
see Figure~\ref{FigContours}.

\begin{figure}

\includegraphics{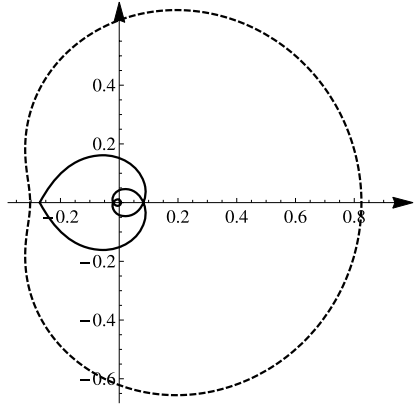}

\caption{(Dashed line) image of $w\mapsto w e^{w}$ for $w\in\Gamma_0$
and (solid line) of $z\mapsto z e^z$ for $z\in\Gamma_-$ (which has
infinitely many small loops around the origin).}
\label{FigContours}
\end{figure}

Computing the finite sum over $k$ leads to
%
\begin{eqnarray}
\label{eq34} %
\qquad\sum_{k=1}^{n_2}
\Psi^{n_1}_{n_1-k}(x_1)\Phi^{n_2}_{n_2-k}(x_2)
&=&\frac{1}{(2\pi\I)^2} \int_{\Gamma_-} \,\D z\oint_{\Gamma_0}\,\D w
\frac
{e^{t z^2/2} e^{-z x_1}(-z)^{n_1}}{e^{t w^2/2} e^{-w x_2} (-w)^{n_2}}
\nonumber
\\[-8pt]
\\[-8pt]
\nonumber
&&{}\times\frac{(1+ w) e^{ w}}{z e^{ z}-w e^{ w}} \biggl(1- \biggl(\frac
{w e^{
w}}{z e^{ z}}
\biggr)^{n_2} \biggr).
\end{eqnarray}
If we do the change of variables $x_i\to x_i- M$ and $n_i\to n_i+M$,
then (\ref{eq34}) becomes
%
\begin{eqnarray}\qquad
\label{eq37} %
 &&\frac{1}{(2\pi\I)^2} \int
_{\Gamma_-} \,\D z\oint_{\Gamma_0}\,\D w \frac
{e^{t z^2/2} e^{-z x_1}(-z)^{n_1}}{e^{t w^2/2} e^{-w x_2}
(-w)^{n_2}}
\frac{(1+ w) e^{ w}}{z e^{ z}-w e^{ w}} e^{ M (z-w)} (z/w)^M
\nonumber
\\[-8pt]
\\[-8pt]
\nonumber
&&{}\qquad-\frac{1}{(2\pi\I)^2} \int_{\Gamma_-} \,\D z\oint_{\Gamma_0}\,\D
w \frac
{e^{t z^2/2} e^{-z x_1}(-z)^{n_1}}{e^{t w^2/2} e^{-w x_2}
(-w)^{n_2}}\frac{(1+ w) e^{ w}}{z e^{ z}-w e^{ w}} \biggl(\frac{w e^{
w}}{z e^{ z}}
\biggr)^{n_2}.
\end{eqnarray}
Denote by $K^{(1)}_t$ the first term in (\ref{eq37}) and by $K^{(2)}_t$
the second term. $K^{(2)}_t$ is independent of $M$.

By Proposition~\ref{propFiniteN} we have
%
\begin{equation}
\Pb \biggl(\bigcap_{k\in S} \bigl\{x_k^{(M)}(t)
\geq a_k\bigr\} \biggr)=\det(\Id-P_a K_{t,M}
P_a)_{L^2(\R\times S)},
\end{equation}
where
%
\begin{eqnarray}
K_{t,M}(x_1,n_1;x_2,n_2)&=&-
\phi ^{(n_1,n_2)}(x_1,x_2)+K^{(1)}_t(x_1,n_1;x_2,n_2)
\nonumber
\\[-8pt]
\\[-8pt]
\nonumber
&&{}+K^{(2)}_t(x_1,n_1;x_2,n_2).
\end{eqnarray}
By Proposition~\ref{propMconv} it follows
%
\begin{equation}
\lim_{M\to\infty}\Pb \biggl(\bigcap_{k\in S}
\bigl\{x_k^{(M)}(t)\geq a_k\bigr\} \biggr) =\Pb
\biggl(\bigcap_{k\in S} \bigl\{x_k(t)\geq
a_k\bigr\} \biggr).
\end{equation}
Therefore to complete the proof we need to show that
%
\begin{equation}
\lim_{M\to\infty} \det(\Id-P_a K_{t,M}
P_a)_{L^2(\R\times S)}= \det\bigl(\Id -P_a
K_t^{\mathrm{flat}} P_a\bigr)_{L^2(\R\times S)}.
\end{equation}
It is easy to verify that
%
\begin{equation}
\bigl\llvert e^{ M (z-w)} (z/w)^M\bigr\rrvert \leq
e^{M  (-1-({1}/2)
|y|+({1}/2)
\ln(4+2|y|+y^2) )} \leq e^{-M/4},
\end{equation}
and to get the bounds
%
\begin{eqnarray}
\bigl|K^{(1)}(x_1,n_1;x_2,n_2)\bigr|&
\leq& C e^{-M/4} e^{(2x_1-x_2)}\nonumber\\
&=& C e^{-M/4}e^{3(x_2-x_1)/2}e^{(x_1+x_2)/2},
\\
\bigl|K^{(2)}(x_1,n_1;x_2,n_2)\bigr|&
\leq& C e^{(2x_1-x_2)} = C e^{3(x_2-x_1)/2}e^{(x_1+x_2)/2} \nonumber
\end{eqnarray}
for some constant $C$ uniform for $x_1,x_2$ bounded from above. Using
the second integral representation in (\ref{eq31}) we get
%
\begin{eqnarray}
\bigl |\phi^{(n_1,n_2)}(x_1,x_2)\bigr|&
\leq &C e^{(x_1-x_2)}\Id(x_1\geq x_2)\Id
(n_2>n_1)
\nonumber
\\[-8pt]
\\[-8pt]
\nonumber
&\leq & C e^{3(x_2-x_1)/2}e^{(x_1+x_2)/2}e^{-|x_1-x_2|/2}\Id(n_2>n_1).
\end{eqnarray}
With these estimates one can show that the Fredholm determinant series
expansion is uniformly integrable/summable in $M$. Dominated
convergence allows us to take the $M\to\infty$ inside the Fredholm
series. The details are exactly as in the proof of Proposition~3.6
of~\cite{BF07}.
This gives
%
\begin{equation}
\lim_{M\to\infty} \det(\Id-P_a K_{t,M}
P_a)_{L^2(\R\times S)} =\det\bigl(\Id -P_a \widetilde
K_t^{\mathrm{flat}} P_a\bigr)_{L^2(\R\times S)},
\end{equation}
where
%
\begin{equation}
\widetilde K_t^{\mathrm{flat}}(x_1,n_1;x_2,n_2)=-
\phi^{(n_1,n_2)}(x_1,x_2) + K^{(2)}(x_1,n_1;x_2,n_2).
\end{equation}

It remains to verify that $\widetilde K_t^{\mathrm{flat}}=K_t^{\mathrm{flat}}$.
Since $K^{(2)}(x_1,n_1;x_2,n_2)$ is given by
%
\begin{equation}
\label{eq44} -\frac{1}{(2\pi\I)^2} \oint_{\Gamma_0}\,\D w \int
_{\Gamma_-} \,\D z\frac
{e^{t z^2/2} e^{-z x_1}(-z)^{n_1}}{e^{t w^2/2} e^{-w x_2}
(-w)^{n_2}}\frac{(1+ w) e^{ w}}{z e^{ z}-w e^{ w}} \biggl(
\frac{w e^{
w}}{z e^{ z}} \biggr)^{n_2},\hspace*{-20pt}
\end{equation}
the pole at $w=0$ is not present.
Let us do the change of variables $W=w e^{ w}$, that is,
%
\begin{equation}
w=w(W)=L_0(W) \qquad\mbox{where }L_0\mbox{ is the
Lambert W function}.
\end{equation}
By the choice of the integration contours, the path for $W$ is still a
simple loop around the origin, and it contains the image of $z e^{ z}$;
see Figure~\ref{FigContours}. Therefore,
%
\begin{eqnarray}
(\ref{eq44})&=&-\frac{1}{(2\pi\I)^2}
\oint_{\Gamma_0}\,\D W \int_{\Gamma
_-} \,\D z\frac{e^{t z^2/2} e^{-z x_1}(-z)^{n_1}}{e^{t w(W)^2/2} e^{-w(W)
x_2} (-w(W))^{n_2}}\nonumber\\
&&{}\times\frac{(W/z e^z)^{n_2}}{z e^{ z}-W}
\\
&=&\frac{1}{2\pi\I} \int_{\Gamma_-} \,\D z\frac{e^{t z^2/2} e^{-z
x_1}(-z)^{n_1}}{e^{t \varphi(z)^2/2} e^{-\varphi(z) x_2}
(-\varphi(z))^{n_2}},\nonumber
\end{eqnarray}
where $\varphi(z)=L_0(z e^{ z})$. At this point, the path $\Gamma_-$
can be deformed to a generic path as in the proposition. The
convergence is ensured by the term $e^{t z^2/2}$.
\end{pf*}

\section{Asymptotic analysis}\label{SectAsympt}
\subsection{Proof of Theorem~\texorpdfstring{\protect\ref{thmAsympFixedTime}}{2.4}}
To ensure convergence of the Fredholm determinants one needs a
pointwise limit as well as integrable bounds of the kernel. The
structure of the proof follows the approach of~\cite{BFP06}. However,
due to the presence of the Lambert function, the search of a steep
descent path is more involved than in previous works.

We will use an explicit expression of the Airy$_1$ kernel defined in
(\ref{defairykernel})
%
\begin{eqnarray}
\label{defAi1K} &&K_{\mathcal{A}_1}(s_1,r_1;s_2,r_2)\nonumber\\
&&\qquad=
-\frac{1}{\sqrt{4\pi(r_2-r_1)}} \exp \biggl(-\frac{(s_2-s_1)^2}{4(r_2-r_1)} \biggr)\Id
(r_2>r_1)
\\
&&\qquad\quad{}+\Ai \bigl(s_1+s_2+(r_2-r_1)^2
\bigr) \exp \biggl((r_2-r_1) (s_1+s_2)+
\frac{2}{3}(r_2-r_1)^3 \biggr).\nonumber
\end{eqnarray}
The scaling limit \eqref{defXt} amounts to setting
%
\begin{eqnarray}
\label{eqscaling} %
n_i&=&-t+2^{5/3}t^{2/3}r_i,
\nonumber
\\[-8pt]
\\[-8pt]
\nonumber
x_i&=&-2^{5/3}t^{2/3}r_i
-(2t)^{1/3}s_i,\qquad i=1,2.
\end{eqnarray}
Finally, we consider a conjugated version of the kernel $K_t^{\mathrm{flat}}$ of Proposition~\ref{propFlat},
%
\begin{equation}
K^{\mathrm{conj}}(x_1,n_1;x_2,n_2)=e^{x_2-x_1}K_t^{\mathrm{flat}}(x_1,n_1;x_2,n_2),
\end{equation}
which decomposes as
%
\begin{equation}
\label{eqKernelDec} K^{\mathrm{conj}}(x_1,n_1;x_2,n_2)=-e^{x_2-x_1}
\phi ^{(n_1,n_2)}(x_1,x_2)+K_0^{\mathrm{conj}}(x_1,n_1;x_2,n_2).\hspace*{-20pt}
\end{equation}

\begin{prop}[(Uniform convergence on compact sets)]\label{propCompConv}
Consider $r_1,r_2\in\R$ as well as $L,\tilde{L} > 0$ fixed.
Then, with $x_i$, $n_i$ defined by \eqref{eqscaling}, the kernel
converges as
%
\begin{equation}
\lim_{t\to\infty}(2t)^{1/3}K^{\mathrm{conj}}(x_1,n_1;x_2,n_2)
\\
=K_{\mathcal
{A}_1}(s_1,r_1;s_2,r_2)
\end{equation}
uniformly for $(s_1, s_2)\in[-L, \tilde{L}]^2$.
\end{prop}

\begin{cor}\label{corFixedD}
Consider \mbox{$r_1,r_2\in\R$} fixed. For any fixed $L,\tilde{L} > 0$
there exists $t_0$ such that for $t>t_0$, the bound
%
\begin{equation}
\bigl\llvert (2t)^{1/3}K^{\mathrm{conj}}(x_1,n_1;x_2,n_2)
\bigr\rrvert \leq\const _{L,\tilde{L}}
\end{equation}
holds for all $s_1,s_2\in[-L,\tilde{L}]$.
\end{cor}

\begin{prop}[(Large deviations)]\label{propLargeD}
For any $L>0$ there exist $\tilde{L}>0$ and $t_0>0$ such that the estimate
%
\begin{equation}
\bigl\llvert (2t)^{1/3}K_0^{\mathrm{conj}}(x_1,n_1;x_2,n_2)
\bigr\rrvert \leq e^{-(s_1+s_2)}
\end{equation}
holds for any $t>t_0$ and $(s_1,s_2)\in[-L,\infty)^2\setminus
[-L,\tilde{L}]^2$.
\end{prop}

\begin{prop}\label{propPhiBound}
For any fixed $r_2-r_1>0$ there exist $\const_1>0$ and $t_0>0$ such
that the bound
%
\begin{equation}
\bigl\llvert (2t)^{1/3}e^{x_2-x_1}\phi^{(n_1,n_2)}(x_1,x_2)
\bigr\rrvert \leq \const _1e^{-|s_2-s_1|}
\end{equation}
holds for any $t>t_0$ and $s_1,s_2\in\R$.
\end{prop}

With these estimates one proves Theorem~\ref{thmAsympFixedTime}.

\begin{pf*}{Proof of Theorem~\ref{thmAsympFixedTime}}
Given the previous bounds, the proof is identical to the proof of
Theorem~2.5 in~\cite{BFP06}. In our case moderate and large deviations
are merged into the single Proposition~\ref{propLargeD}. The constants
appearing in~\cite{BFP06} specialize to $\kappa=2^{1/3}$ and $\mu
=-2^{5/3}$ in our setting.
\end{pf*}

Now let us prove the convergence of the kernel.

\begin{pf*}{Proof of Proposition~\ref{propCompConv}}
We start with the first part of the conjugated kernel \eqref
{eqKernelDec} in its integral representation \eqref{eq31},
%
\begin{equation}
\label{eq62} (2t)^{1/3}e^{x_2-x_1}\phi^{(n_1,n_2)}(x_1,x_2)
= \frac
{(2t)^{1/3}}{2\pi
\I} \int_{\I\R-\delta}\,\D z \frac{e^{(z+1)(x_2-x_1)}}{(-z)^{n_2-n_1}}.
\end{equation}
Setting $\delta=1$ and using the change of variables
$z=-1+(2t)^{-1/3}\zeta$ as well as the shorthand $r=r_2-r_1$ and
$s=s_2-s_1$, we have
%
\begin{equation}\quad
\eqref{eq62}= \frac{1}{2\pi\I} \int_{\I\R}\,\D\zeta
\frac
{e^{(2t)^{-1/3}\zeta(x_2-x_1)}}{(1-(2t)^{-1/3}\zeta)^{n_2-n_1}}= \frac{1}{2\pi\I} \int_{\I\R}\,\D\zeta
e^{-s\zeta}f_t(\zeta,r)
\end{equation}
with
%
\begin{equation}
f_t(\zeta,r)=\frac{e^{-2^{4/3}t^{1/3}r\zeta}}{(1-(2t)^{-1/3}\zeta
)^{2^{5/3}t^{2/3}r}}=e^{-2^{4/3}t^{1/3}r\zeta-2^{5/3}t^{2/3}r\log
(1-(2t)^{-1/3}\zeta)}.\hspace*{-33pt}
\end{equation}
Since this integral is $0$ for $r\leq0$ we can assume $r>0$ from now
on. The function $f_t(\zeta,r)$ satisfies the pointwise limit $\lim_{t\to\infty}f_t(\zeta,r)=e^{r\zeta^2}$. Applying Bernoulli's
inequality, we arrive at the $t$-independent integrable bound
%
\begin{eqnarray}
\qquad \bigl|f_t(\zeta,r)\bigr|&=&\bigl|1-(2t)^{-1/3}
\zeta\bigr|^{-2^{5/3}t^{2/3}r}= \bigl(1+(2t)^{-2/3}|\zeta|^2
\bigr)^{-2^{2/3}t^{2/3}r}
\nonumber
\\[-8pt]
\\[-8pt]
\nonumber
&\leq&\bigl(1+r|\zeta|^2\bigr)^{-1}.
\end{eqnarray}
Thus by dominated convergence
%
\begin{eqnarray}
&&\biggl\llvert \frac{1}{2\pi\I} \int_{\I\R}\,\D\zeta
\bigl(e^{-s\zeta
}f_t(\zeta,r)-e^{-s\zeta+r\zeta^2} \bigr)\biggr
\rrvert
\nonumber
\\[-8pt]
\\[-8pt]
\nonumber
&&\qquad\leq\frac{1}{2\pi} \int_{\I\R
}|\D\zeta|
\bigl|f_t(\zeta,r)-e^{r\zeta^2} \bigr|\stackrel{t\to \infty } {
\longrightarrow}0.
\end{eqnarray}
This implies that the convergence of the integral is uniform in $s$.
The limit is easily identified as
%
\begin{eqnarray}\qquad
\lim_{t\to\infty}-(2t)^{1/3}e^{x_2-x_1}
\phi ^{(n_1,n_2)}(x_1,x_2)&=&-\frac{1}{2\pi\I} \int
_{\I\R}\,\D\zeta \,e^{-s\zeta
+r\zeta^2}\Id(r>0)
\nonumber
\\[-8pt]
\\[-8pt]
\nonumber
&=& -\frac{1}{\sqrt{4\pi r}}e^{-s^2/4r}\Id(r>0),
\end{eqnarray}
which is the first part of the kernel \eqref{defAi1K}.

Now we turn to the main part of the kernel,
%
\begin{equation}
K_0^{\mathrm{conj}}(x_1,n_1;x_2,n_2)=
\frac{1}{2\pi\I} \int_{\Gamma_-} \,\D z\frac{e^{t z^2/2} e^{-(z+1)x_1}(-z)^{n_1}}{e^{t \varphi(z)^2/2}
e^{-(\varphi(z)+1)x_2} (-\varphi(z))^{n_2}}.\hspace*{-20pt}
\end{equation}
Inserting the scaling \eqref{eqscaling} and using the identity
$z/\varphi(z)=e^{\varphi(z)-z}$ we define the functions
%
\begin{eqnarray}
 f_3(z)&=&\tfrac{1}{2}
\bigl(z^2+2z -\varphi(z)^2-2\varphi(z) \bigr),\nonumber
\\
f_2(z)&=&2^{5/3} \bigl(r_1 \bigl[z+1+\log (-z )
\bigr]-r_2 \bigl[\varphi(z)+1+\log \bigl(-\varphi(z) \bigr) \bigr]
\bigr),
\\
f_1(z)&=&2^{1/3} \bigl(s_1(z+1)-s_2
\bigl(\varphi(z)+1\bigr) \bigr),\nonumber
\end{eqnarray}
which transforms the kernel to
%
\begin{equation}\qquad
K_0^{\mathrm{conj}}(x_1,n_1;x_2,n_2)=
\frac{1}{2\pi\I} \int_{\Gamma_-} \,\D z \exp \bigl(tf_3(z)+t^{2/3}f_2(z)+t^{1/3}f_1(z)
\bigr).\hspace*{-15pt}
\end{equation}
Define for $0\leq\rho<1$ a contour by
%
\begin{equation}
\Gamma^\rho= \bigl\{L_{\lfloor\tau\rfloor} \bigl(-(1-\rho)e^{2\pi
\I\tau
-1}
\bigr),\tau\in\R\setminus[0,1) \bigr\}
\end{equation}
with $L_k(z)$ being the $k$th branch of the Lambert W function. We
specify the contour $\Gamma_-$ by $\Gamma:=\Gamma^0$, which is shown in
Figure~\ref{figGamma}, along with a $\rho$-deformed version, which will
be used later in the asymptotic analysis.
Lemma~\ref{lemContour} ensures that this contour is an admissible
choice. By Lemma~\ref{lemf3} (with $\rho=0$), $\Gamma$ is a steep
descent curve for the function $f_3$ with maximum real part $0$ at
$z=-1$ and strictly negative everywhere else. We can therefore restrict
the contour to \mbox{$\Gamma_{\delta}=\{z\in\Gamma,|z+1|<\delta\}
$} by
making an error which is exponentially small in $t$, uniformly for
$s_i\in[-L,\tilde{L}]$.

\begin{figure}

\includegraphics{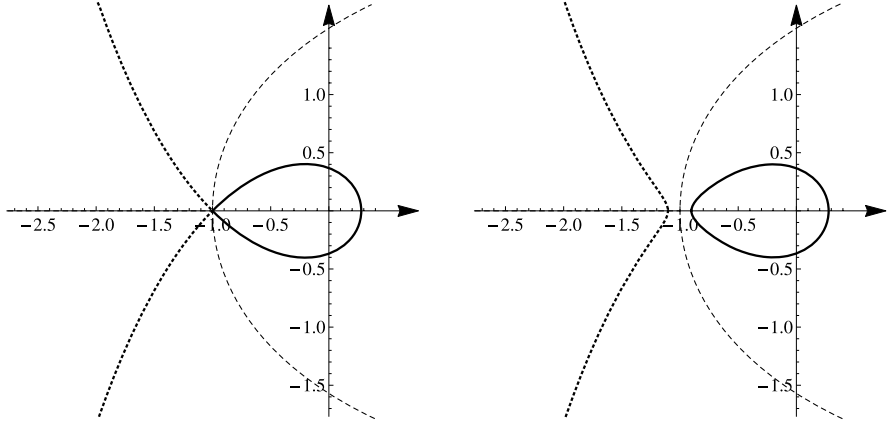}

\caption{(Dotted line) the contour $\Gamma^\rho$ and (solid line)
its image under $\varphi$ for (left picture) $\rho=0$ and (right
picture)
some small positive $\rho$. The dashed lines separate the ranges
of the principal branch $0$ (right) and the branches $1$ (top left)
and $-1$ (bottom left).}\vspace*{-3pt}
\label{figGamma}
\end{figure}

By (4.22) in~\cite{CGHJK96} the Lambert W function can be expanded
around the branching point $-e^{-1}$ as
%
\begin{equation}
\label{LambTayl} L_0(z)=-1+p-\tfrac{1}{3}p^2+
\tfrac{11}{72}p^3+\cdots,
\end{equation}
with $p(z)=\sqrt{2(ez+1)}$. Inserting the Taylor series of $ze^z$
provides the expansion of $\varphi$ and hence of the functions $f_i$ in
the neighbourhood of $z=-1$,
%
\begin{eqnarray}
 \varphi(-1+\zeta)&=&-1-\zeta-\tfrac{2}{3}
\zeta^2+\Or\bigl(\zeta^3\bigr),\nonumber
\\
f_3(-1+\zeta)&=&-\tfrac{2}{3}\zeta^3+\Or\bigl(
\zeta^4\bigr),
\nonumber
\\[-10pt]
\\[-10pt]
\nonumber
f_2(-1+\zeta )&=&2^{2/3}r\zeta^2+\Or\bigl(
\zeta^3\bigr),
\\
f_1(-1+\zeta)&=&2^{1/3}(s_1+s_2)
\zeta+\Or\bigl(\zeta^2\bigr). \nonumber
\end{eqnarray}

The $\Or$-terms should be understood as uniform in $s_i$ for $s_i\in
[-L,\tilde{L}]$ and with $r_i$ fixed. Let $\tilde{f}_i(\zeta)$ be the
expression $f_i(\zeta)$ omitting the error term. Define also
\[
F(\zeta)= \exp \bigl(tf_3(-1+\zeta)+t^{2/3}f_2(-1+
\zeta )+t^{1/3}f_1(-1+\zeta) \bigr)
\]
and the corresponding version $\tilde{F}(\zeta)$ without the errors.
Let further $\widehat{\Gamma}_\delta= \{z+1; z\in\Gamma
_\delta \}$. Using the inequality $\llvert e^x-1\rrvert \leq
\llvert x\rrvert e^{\llvert x\rrvert }$, the error made by integrating over
$\tilde
{F}$ instead of $F$ can be estimated as
%
\begin{eqnarray}
&&\biggl\llvert \frac{(2t)^{1/3}}{2\pi\I} \int
_{\widehat{\Gamma}_\delta
}\,\mathrm{d}\zeta \bigl(F(\zeta)-\tilde{F}(\zeta) \bigr)\biggr
\rrvert\nonumber
\\
&&\qquad\leq\frac{(2t)^{1/3}}{2\pi}\int_{\widehat{\Gamma}_\delta
}\,\mathrm{d}\zeta \bigl|\tilde{F}(\zeta)
\bigr|e^{\Or (\zeta^4t+\zeta^3t^{2/3}+\zeta
^2t^{1/3}+\zeta )}\nonumber\\
&&\qquad\quad{}\times\Or \bigl(\zeta^4t+\zeta^3t^{2/3}+
\zeta ^2t^{1/3}+\zeta \bigr)
\\
&&\qquad\leq\frac{(2t)^{1/3}}{2\pi}\int_{\widehat{\Gamma}_\delta}\,\mathrm{d}\zeta \bigl\llvert
e^{t\tilde{f}_3(\zeta-1)(1+\chi_3)+t^{2/3}\tilde{f}_2(\zeta
-1)(1+\chi_2)+t^{1/3}\tilde{f}_1(\zeta-1)(1+\chi_1)}\bigr\rrvert\nonumber
\\
&&\qquad\quad{} \times\Or \bigl(\zeta^4t+\zeta^3t^{2/3}+
\zeta ^2t^{1/3}+\zeta \bigr), \nonumber
\end{eqnarray}
where $\chi_1$, $\chi_2$, $\chi_3$ are constants, which can be made as
small as desired for $\delta$ small enough. Since the contour
$\widehat
{\Gamma}_\delta$ is close to $\{\llvert \zeta\rrvert  e^{
({3}/{4})\I\pi
\sgn(\zeta) },\zeta\in(-\delta,\delta)\}$ the leading term in the
exponential, \mbox{$t\tilde{f}_3(\zeta-1)(1+\chi_3)=-\frac
{2}{3}\zeta
^3(1+\chi_3)t$} has negative real part and therefore ensures the
integral to stay bounded for $t\to\infty$. By the change of variables
$\zeta=t^{-1/3}\xi$ the $t^{1/3}$ prefactor cancels and the remaining
$\Or$-terms imply that the overall error is $\Or(t^{-1/3})$.

The final step is to evaluate $\frac{(2t)^{1/3}}{2\pi\I} \int_{\widehat
{\Gamma}_\delta}\,\mathrm{d}\zeta \tilde{F}(\zeta)$. The change of variables
\mbox{$\zeta=-(2t)^{-1/3}\xi$} converts the contour of integration to
\mbox{$\eta_t=\{-(2t)^{1/3}\zeta,\zeta\in\widehat{\Gamma}_\delta
\}$},
and hence
%
\begin{eqnarray}\qquad
\frac{(2t)^{1/3}}{2\pi\I} \int_{\widehat{\Gamma}_\delta}\,\mathrm{d}
\zeta \tilde {F}(\zeta) &=& \frac{(2t)^{1/3}}{2\pi\I} \int_{\widehat{\Gamma
}_\delta
}\,\mathrm{d}\zeta
\,e^{-({2}/{3})t\zeta^3+r(2t)^{2/3}\zeta
^2+(s_1+s_2)(2t)^{1/3}\zeta}
\nonumber
\\[-8pt]
\\[-8pt]
\nonumber
&=& \frac{-1}{2\pi\I}\int_{\eta_t}\,\mathrm{d}\xi \,e^{({\xi^3}/{3})+r\xi
^2-(s_1+s_2)\xi}.
\end{eqnarray}
For $t\to\infty$ the contour $\eta_t$ converges to $\eta_\infty=\{
\llvert \xi\rrvert  e^{({\I\pi}/{4})\sgn(\xi) },\xi\in\R\}$. Since
there are
no poles in the relevant region with the cubic term guaranteeing
convergence, we can change $\eta_\infty$ to the usual Airy contour
$\eta
=\{\llvert \xi\rrvert  e^{({\I\pi}/{3})\sgn(\xi) },\break \xi\in\R\}
$, so that
%
\begin{eqnarray}
\lim_{t\to\infty}\frac{(2t)^{1/3}}{2\pi\I} \int
_{\widehat
{\Gamma}_\delta
}\,\mathrm{d}\zeta \tilde{F}(\zeta) &=&\frac{-1}{2\pi\I}\int
_{\eta}\,\mathrm{d}\xi \,e^{( {\xi^3}/{3})+r\xi
^2-(s_1+s_2)\xi}
\nonumber
\\[-8pt]
\\[-8pt]
\nonumber
&=&\Ai \bigl(s_1+s_2+r^2
\bigr)e^{({2}/{3})r^3+(s_1+s_2)r}.
\end{eqnarray}
\upqed\end{pf*}

\subsection{Kernel bounds}
\subsubsection*{Bound on the main part of the kernel}

\mbox{}
\begin{pf*}{Proof of Proposition~\ref{propLargeD}}
The result for $(s_1,s_2)\in[-L,\tilde L]^2$ follows from the
estimates in the proof of Proposition~\ref{propCompConv}. Thus let us
consider the region $(s_1,s_2)\in[-L,\infty]^2\setminus[-L,\tilde
{L}]^2$, so the inequality $s_1+s_2\geq\tilde{L}-L\geq0$ holds. Define
also nonnegative variables $\tilde{s}_i=s_i+L$.
Since $\tilde{s}_i$ are no longer bounded from above, we slightly
redefine our functions $f$ by decomposing $f_1=f_{11}+f_{12}$,
%
\begin{eqnarray}
\qquad f_3(z)&=&\tfrac{1}{2}
\bigl(z^2+2z -\varphi(z)^2-2\varphi(z) \bigr),\nonumber
\\
f_2(z)&=&2^{5/3} \bigl(r_1 \bigl[z+1+\log (-z )
\bigr]-r_2 \bigl[\varphi(z)+1+\log \bigl(-\varphi(z) \bigr) \bigr]
\bigr),
\nonumber
\\[-8pt]
\\[-8pt]
\nonumber
f_{11}(z)&=&2^{1/3} \bigl(\tilde{s}_1(z+1)-
\tilde{s}_2\bigl(\varphi (z)+1\bigr) \bigr),
\\
f_{12}(z)&=&2^{1/3}L \bigl(\varphi(z)-z \bigr). \nonumber
\end{eqnarray}
Using the shorthand $G(z)=tf_3(z)+t^{2/3}f_2(z)+t^{1/3}
(f_{11}(z)+f_{12}(z) )$ the kernel that we want to bound attains
the form
%
\begin{equation}
(2t)^{1/3}K_0^{\mathrm{conj}}(x_1,n_1;x_2,n_2)=
\frac{(2t)^{1/3}}{2\pi\I} \int_{\Gamma} \,\D z \,e^{G(z)}.
\end{equation}
We deform the contour $\Gamma$ to $\Gamma^\rho=\{L_{\lfloor\tau
\rfloor
}(-e^{2\pi\I\tau-1}(1-\rho)),\tau\in\R\setminus[0,1)\}$, where
$\rho$
is given by
%
\begin{equation}
\label{eqrho} \rho=2^{-5/3}\min \bigl\{t^{-2/3}(s_1+s_2),
\e \bigr\}
\end{equation}
for some small $\e>0$ to be chosen in the following. The point where
$\Gamma^\rho$ crosses the real line is given by $z_0=-1-\sqrt{2\rho
}+\Or
(\rho)$ according to Lemma~\ref{lemContour}. We also decompose the
kernel as
%
\begin{equation}
(2t)^{1/3}K_0^{\mathrm{conj}}(x_1,n_1;x_2,n_2)=e^{G(z_0)}
\frac
{(2t)^{1/3}}{2\pi\I}\int_{\Gamma^\rho} \,\D z \,e^{G(z)-G(z_0)}.
\end{equation}

For estimating the first factor one uses the fact that $\rho<\e$ and
applies the Taylor approximation,
%
\begin{eqnarray}
 f_3(-1+\zeta)&=&-\tfrac{2}{3}
\zeta^3+\Or\bigl(\zeta^4\bigr),\nonumber
\\
f_2(-1+\zeta)&=&2^{2/3}r\zeta^2+\Or\bigl(
\zeta^3\bigr),
\nonumber
\\[-8pt]
\\[-8pt]
\nonumber
f_{11}(-1+\zeta)&=&2^{1/3}(\tilde{s}_1+
\tilde{s}_2)\zeta+\Or\bigl(\zeta ^2\bigr),
\\
f_{12}(-1+\zeta)&=&-2^{4/3}L\zeta+\Or\bigl(
\zeta^2\bigr).\nonumber
\end{eqnarray}
Inserting $\zeta= -\sqrt{2\rho}+\Or(\rho)$ and using the two
inequalities for $\rho$ coming from~\eqref{eqrho}, we can bound the
arguments of the exponential as
%
\begin{eqnarray}
\Re\bigl(tf_3(-1+\zeta)\bigr)&\leq&
\tfrac{2}{3}t(2\rho)^{3/2} \bigl(1+\Or (\sqrt{\rho }) \bigr)\nonumber\\
&\leq&
\tfrac{1}{3}(s_1+s_2)^{3/2} \bigl(1+\Or(
\sqrt{\e }) \bigr),\nonumber
\\
\Re\bigl(t^{2/3}f_2(-1+\zeta)\bigr)&\leq&|r|(s_1+s_2)
\bigl(1+\Or(\sqrt{\e }) \bigr),
\nonumber
\\[-8pt]
\\[-8pt]
\nonumber
\Re\bigl(t^{1/3}f_{11}(-1+\zeta)\bigr)&\leq&-(
\tilde{s}_1+\tilde {s}_2) (s_1+s_2)^{1/2}
\bigl(1+\Or(\sqrt{\e}) \bigr)
\\
&\leq&-(s_1+s_2)^{3/2} \bigl(1+\Or(\sqrt{\e})
\bigr),\nonumber
\\
\Re\bigl(t^{1/3}f_{12}(-1+\zeta)\bigr)&\leq&2L(s_1+s_2)^{1/2}
\bigl(1+\Or (\sqrt{\e }) \bigr). \nonumber
\end{eqnarray}
Now choose first $\e$ such that the $f_{11}$ term dominates the $f_3$
term. Then choose $\tilde{L}$ such that the $(s_1+s_2)^{3/2}$-terms
dominate all other terms, leading to the bound
%
\begin{equation}
\label{eqG0bound} \bigl\llvert e^{G(z_0)}\bigr\rrvert \leq e^{-\const_2(s_1+s_2)^{3/2}}
\end{equation}
for some $\const_2>0$.

The remaining task is to show boundedness of the integral
$(2t)^{1/3}\times \int_{\Gamma^\rho} \,\D z \,e^{G(z)-G(z_0)}$. At first we
notice that by Lemma~\ref{lemContour} the terms $z+1$ and $-
(\varphi(z)+1 )$ attain their maximum real part at $z_0$, so
$\tilde{s}_i\geq0$ results in
%
\begin{equation}
\Re \bigl(f_{11}(z)-f_{11}(z_0) \bigr)\leq0
\end{equation}
along $\Gamma^\rho$. This leads to the estimate
%
\begin{equation}
\label{eq77} \biggl\llvert \int_{\Gamma^\rho} \,\D z
\,e^{G(z)-G(z_0)}\biggr\rrvert \leq\int_{\Gamma
^\rho} |\D z|
\bigl|e^{t\hat{f}_3(z)+t^{2/3}\hat{f}_2(z)+t^{1/3}\hat
{f}_{12}(z)}\bigr |,
\end{equation}
where $\hat{f}_i(z)=f_i(z)-f_i(z_0)$. Notice that in the integral on
the right-hand side the variables $s_i$ no longer appear. Integrability
is ensured by Lemmas~\ref{lemContour}, \ref{lemf3} and claim (5),
respectively. As $\Gamma^\rho$ is a steep descent path for $\hat{f}_3$
by Lemma~\ref{lemf3}, we can restrict the contour to a $\delta
$-neighborhood of the critical point, $\Gamma^\rho_\delta=\{
z\in
\Gamma^\rho,  |z-z_0|<\delta\}$, at the expense of an error of order
$\Or(e^{-\const_\delta t})$.

Since the contour $\Gamma^\rho_\delta$ approaches a straight vertical
line, we can set $z=z_0+\I\xi$ and expand for small $\xi$ as
%
\begin{eqnarray}
\Re\bigl(\hat{f}_3(z_0+\I
\xi)\bigr)&=&-2\sqrt{2\rho} \xi^2 \bigl(1+\Or(\xi ) \bigr) \bigl(1+\Or(
\sqrt{\rho}) \bigr),\nonumber
\\
\Re\bigl(\hat{f}_2(z_0+\I\xi)\bigr)&=&-2^{2/3}r
\xi^2 \bigl(1+\Or(\xi) \bigr) \bigl(1+\Or(\sqrt{\rho}) \bigr),
\\
\Re\bigl(\hat{f}_{12}(z_0+\I\xi)\bigr)&=&
\tfrac{1}{3}2^{4/3}L\xi^2 \bigl(1+\Or(\xi ) \bigr)
\bigl(1+\Or(\sqrt{\rho}) \bigr). \nonumber
\end{eqnarray}
By choosing $\delta$ and $\e$ small enough there are some constants
$\chi_3,\chi_2,\chi_1$ close to $1$ such that
%
\begin{eqnarray}\quad
\label{eq78} \int_{\Gamma^\rho_\delta} \,\D z
\bigl\llvert e^{G(z)-G(z_0)}\bigr\rrvert &\leq& \int_{-\delta}^\delta
\,\D\xi \,e^{\xi^2 (-\chi_3 2\sqrt{2\rho
}t-\chi_2
2^{2/3}rt^{2/3}+\chi_1({2^{4/3}}/{3})Lt^{1/3} )}
\nonumber
\\[-8pt]
\\[-8pt]
\nonumber
&=&\int_{-\delta}^\delta\,\D\xi \,e^{\eta t^{2/3}\xi^2}\leq
t^{-1/3}\sqrt{\frac{\pi}{\eta}},
\end{eqnarray}
where
%
\begin{equation}
\eta=2\sqrt{2\rho}t^{1/3}\chi_3+2^{2/3}r
\chi_2-\tfrac
{2^{4/3}}{3}Lt^{-1/3}\chi_1.
\end{equation}
Since $\sqrt{2\rho}t^{1/3}\geq2^{1/3}\min \{\sqrt{\tilde
{L}-L},\sqrt
{\e}t^{1/3} \}$, the first term dominates the other two for
$\tilde
{L}$ and $t$ large enough. Then $\eta$ is bounded from below by some
positive constant $\eta_0$. Combining \eqref{eqG0bound} and \eqref
{eq78} we finally arrive at
%
\begin{eqnarray}\qquad
&&\bigl |(2t)^{1/3}K^{\mathrm{conj}}(x_1,n_1;x_2,n_2)
\bigr|
\nonumber
\\[-8pt]
\\[-8pt]
\nonumber
&&\qquad\leq\frac
{(2t)^{1/3}}{2\pi}t^{-1/3}\sqrt{\frac{\pi}{\eta_0}}e^{-\const
_2(s_1+s_2)^{3/2}}
\bigl(1+\Or\bigl(e^{-c(\delta)t}\bigr) \bigr) \leq e^{-(s_1+s_2)},
\end{eqnarray}
where the last inequality holds for $t$ and $\tilde{L}$ large enough.
\end{pf*}

\subsubsection*{Bound on $\phi$}

\mbox{}
\begin{pf*}{Proof of Proposition~\ref{propPhiBound}}
We start with the elementary representation of $\phi$ given in \eqref
{eq30} and insert the scaling
%
\begin{eqnarray}
\label{phiBound1} %
&&(2t)^{1/3}e^{x_2-x_1}
\frac{(x_1-x_2)^{n_2-n_1-1}}{(n_2-n_1-1)!}
\nonumber\\
&&\qquad=(2t)^{1/3}e^{-2^{5/3}t^{2/3}r-(2t)^{1/3}s}\frac
{(2^{5/3}t^{2/3}r+(2t)^{1/3}s)^{2^{5/3}t^{2/3}r-1}}{(2^{5/3}t^{2/3}r-1)!}
\nonumber
\\[-8pt]
\\[-8pt]
\nonumber
&&\qquad=\bigl(1+\Or\bigl(t^{-2/3}\bigr)\bigr)\frac{2^{1/3}}{\sqrt{2\pi r}}
\frac
{e^{-(2t)^{1/3}s}}{1+2^{-4/3}t^{-1/3}s/r}\\
&&\qquad\quad{}\times\bigl(1+2^{-4/3}t^{-1/3}s/r
\bigr)^{2^{5/3}t^{2/3}r}. \nonumber
\end{eqnarray}
Since the factorial depends on $r$ and $t$ only, the error from the
Stirling formula is uniform in $s$. Introducing $\tilde
{s}=2^{-4/3}t^{-1/3}s/r$ and some $\const_3$ depending only on $r$ we have
%
\begin{equation}
\label{phiBound2} \bigl\llvert \eqref{phiBound1}\bigr\rrvert \leq\const_3
e^{-(2t)^{1/3}s}(1+\tilde {s})^{2^{5/3}t^{2/3}r-1}
\end{equation}
for $t$ large enough.

Applying the inequality $1+x\leq\exp (x-x^2/2+x^3/3 )$ we
arrive at
%
\begin{eqnarray}
 \bigl\llvert \eqref{phiBound1}\bigr\rrvert &\leq&
\const_3 e^{-(2t)^{1/3}s+
(2^{5/3}t^{2/3}r-1 ) (\tilde{s}-\tilde{s}^2/2+\tilde
{s}^3/3 )}
\nonumber
\\[-8pt]
\\[-8pt]
\nonumber
&=& \const_3 e^{-({s^2}/{(4r)}) (1-({2}/{3})\tilde{s}
)- (\tilde{s}-\tilde{s}^2/2+\tilde{s}^3/3 )}.
\end{eqnarray}
In the case $|\tilde{s}|\leq1$ we now use the basic inequality
%
\begin{equation}
e^{-a^2/b}\leq e^be^{-|a|}
\end{equation}
to obtain the desired bound.

Inserting the scaling into the conditions $n_2>n_1$ and $x_1\geq x_2$
appearing in \eqref{eq30} results in $\tilde{s}\geq-1$. So we are left
to prove the claim for $\tilde{s}>1$.

From \eqref{phiBound2} one obtains
%
\begin{eqnarray}
\bigl\llvert \eqref{phiBound1}\bigr\rrvert &\leq&\const_3 \tfrac{1}{2}e^{-\tilde
{s}\cdot2^{5/3}t^{2/3}r}(1+
\tilde{s})^{2^{5/3}t^{2/3}r}
\nonumber
\\[-8pt]
\\[-8pt]
\nonumber
& =&\const_3 \tfrac{1}{2} \bigl((1+
\tilde{s})e^{-\tilde{s}} \bigr)^{2^{5/3}t^{2/3}r}.
\end{eqnarray}
The elementary estimate $(1+\tilde{s})e^{-\tilde{s}}\leq e^{-\tilde
{s}/4}$ finally results in
%
\begin{eqnarray}
\bigl\llvert \eqref{phiBound1}\bigr\rrvert &\leq&\const_3 \tfrac{1}{2}
e^{-({1}/{4})\tilde{s}\cdot2^{5/3}t^{2/3}r}
\nonumber
\\[-8pt]
\\[-8pt]
\nonumber
&=&\const_3 \tfrac{1}{2}e^{-({1}/{4})(2t)^{1/3}s}\leq
\const_1 e^{-s}
\end{eqnarray}
for $t\geq32$.
\end{pf*}

\section{Tagged particle and slow decorrelations}\label{SectTagged}
In this section we want to prove the following result and then use it
to show Theorem~\ref{thmTagged}.

\begin{thmm}\label{ThmAsymptSpaceTime}
Let us fix a $\nu\in[0,1)$, choose any $\theta_1,\ldots,\theta
_m\in
[-t^{\nu},t^{\nu}]$, $u_1,\ldots,u_m\in\R$ and define the rescaled
random variables
%
\begin{equation}
\label{eq5.1} X_t^{\mathrm{resc}}(u_k,
\theta_k):=-\frac{x_{[-t+u_k 2^{5/3}
t^{2/3}+\theta
_k]}(t+\theta_k)+2\theta_k+u_k 2^{5/3} t^{2/3}}{(2t)^{1/3}}.
\end{equation}
Then, for any $s_1,\ldots,s_m\in\R$ fixed, it holds
%
\begin{equation}
\lim_{t\to\infty}\Pb \Biggl(\bigcap_{k=1}^m
\bigl\{X_t^{\mathrm{resc}}(u_k,\theta_k)\leq
s_k \bigr\} \Biggr) = \Pb \Biggl(\bigcap
_{k=1}^m \bigl\{{\cal A}_1(u_k)
\leq s_k \bigr\} \Biggr).
\end{equation}
\end{thmm}

As a corollary we have Theorem~\ref{thmTagged}.

\begin{pf*}{Proof of Theorem~\ref{thmTagged}}
This follows by taking $\theta_k=\tau_k 2^{5/3}t^{2/3}$ and
$u_k=-\tau
_k$ in Theorem~\ref{ThmAsymptSpaceTime}. Indeed,
%
\begin{equation}
X_t^{\mathrm{resc}}(u_k,\theta_k)=-
\frac{x_{[-t]}(t+\tau_k
2^{5/3}t^{2/3})+\tau_k 2^{5/3} t^{2/3}}{(2t)^{1/3}},
\end{equation}
which by translation invariance by an integer has the same distribution
as $X^{\mathrm{tagged}}_t(\tau_k)$ [the difference due to the integer value
approximation is at most $1/(2t)^{1/3}$, which is asymptotically irrelevant].
\end{pf*}

For the proof of Theorem~\ref{ThmAsymptSpaceTime} we need the following
slow-decorrelation result.

\begin{prop}\label{PropSlowDecorrelation}
For a $\nu\in[0,1)$, let us consider $\theta\in[-t^\nu,t^\nu]$.
Then, for any $\e>0$,
%
\begin{equation}
\lim_{t\to\infty}\Pb \bigl( \bigl|x_{n+\theta}(t+\theta
)-x_n(t)+2\theta\bigr|\geq \e t^{1/3} \bigr)=0.
\end{equation}
\end{prop}

\begin{pf}
Without loss of generality we consider $\theta\geq0$. For $\theta<0$
one just has to denote $\tilde t=t+\theta$ so that $\tilde t-\theta=t$,
and the proof remains valid with $t$ replaced by $\tilde t$. Recall
that by definition we have
%
\begin{equation}
\label{eqSlow3} x_m(t)=-\max_{k\leq m}\bigl
\{Y_{k,m}(t)-x_k(0)\bigr\}, \qquad 1\leq m \leq N,
\end{equation}
with $x_k(0)=-k$. We also define
%
\begin{equation}
x_m^{\mathrm{step}}(t)=-\max_{1\leq k\leq m}\bigl
\{Y_{k,m}(t)\bigr\} =-Y_{1,m}(t),\qquad 1\leq m \leq N.
\end{equation}
First we need an inequality, namely
%
\begin{eqnarray}
\label{eqCoupling} %
 -x_{n+\theta}(t+\theta)&=&\max
_{k\leq n+\theta}\bigl\{k+Y_{k,n+\theta
}(t+\theta)\bigr\}\geq\max
_{k\leq n}\bigl\{k+Y_{k,n+\theta}(t+\theta)\bigr\}
\nonumber\\
&=&\max_{k\leq n}\Biggl\{k+\sup_{0\leq s_{k+1}\leq\cdots\leq s_{n+\theta
+1}=t+\theta}\sum
_{i=k}^{n+\theta} \bigl(B_i(s_{i+1})-B_i(s_i)
\bigr)\Biggr\}
\nonumber
\\[-8pt]
\\[-8pt]
\nonumber
&\geq&\max_{k\leq n}\Biggl\{k+\mathop{\sup_{
0\leq s_{k+1}\leq\cdots\leq s_{n+\theta+1}=t+\theta}}_{\mathrm{with\ }
s_{n+1}=t
}\sum
_{i=k}^{n+\theta} \bigl(B_i(s_{i+1})-B_i(s_i)
\bigr)\Biggr\}
\\
&=&-x_n(t) - \widetilde{x}_\theta^{\mathrm{step}}(\theta),
\nonumber
\end{eqnarray}
with
%
\begin{equation}
\widetilde{x}_\theta^{\mathrm{step}}(\theta)=\sup_{t\leq
s_{n+2}\leq
\cdots\leq s_{n+\theta}\leq t+\theta}
\sum_{i=n+1}^{n+\theta} \bigl(B_i(s_{i+1})-B_i(s_i)
\bigr).
\end{equation}
Remark that $x_n(t)$ and $\widetilde{x}_\theta^{\mathrm
{step}}(\theta)$
are independent and $\widetilde{x}_\theta^{\mathrm{step}}(\theta
)\stackrel{d}{=}x_\theta^{\mathrm{step}}(\theta)$.

From Theorem~\ref{thmAsympFixedTime} we have
%
\begin{eqnarray}
\label{eq5.6} %
 \chi_1(t)&:=&-
\frac{x_n(t)+n+t}{(2t)^{1/3}}\stackrel {D} {\Longrightarrow }D_1,
\nonumber
\\[-8pt]
\\[-8pt]
\nonumber
\chi_2(t)&:=&-\frac{x_{n+\theta}(t+\theta)+n+t+2\theta
}{(2t)^{1/3}}\stackrel{D} {\Longrightarrow}D_1,
\end{eqnarray}
with $D_1(s)=F_1(2s)$, $F_1$ being the GOE Tracy--Widom distribution
function~\cite{TW96}.
Further, it is known by the connection with the GUE random
matrices~\cite{Bar01,TW94}, that
%
\begin{equation}
\label{eq5.10} -\frac{\widetilde x^{\mathrm{step}}_\theta(\theta)+2\theta}{(2\theta
)^{1/3}}\stackrel{D} {\Longrightarrow}D_2,
\end{equation}
where $D_2(s)=F_2(2^{1/3} s)$, $F_2$ being the GUE Tracy--Widom
distribution function~\cite{TW94}.
Therefore,
%
\begin{equation}
\label{eq5.8} \chi_3(t):=-\frac{\widetilde x^{\mathrm{step}}_\theta(\theta)+2\theta
}{(2t)^{1/3}}\stackrel{D} {
\Longrightarrow}0,
\end{equation}
by (\ref{eq5.10}) and $\theta/t\to0$. By (\ref{eq5.6}) and (\ref
{eq5.8}) we have $\chi_1(t)+\chi_3(t)\stackrel{D}{\Longrightarrow}D_1$.
Further, (\ref{eqCoupling}) implies that
%
\begin{equation}
\chi_2(t)=\chi_1(t)+\chi_3(t)+R_t
\end{equation}
for some random variable $R_t\geq0$. Since both $\chi_2(t)$ and $\chi
_1(t)+\chi_3(t)$ converges in distribution to $D_1$ and $R_t\geq0$, by
Lemma~4.1 of~\cite{BC09} (reported below), we have $R_t\to0$ in
probability as $t\to\infty$. This together with (\ref{eq5.8}) leads to
$\chi_2(t)-\chi_1(t)\to0$ in probability, which is the rescaled
version of our statement.
\end{pf}

\begin{lem}[(Lemma~4.1 of~\cite{BC09})]\label{BAC_lemma}
Consider two sequences of random variables $\{X_n\}$ and $\{\tilde
{X}_n\}$ such that for each $n$,
$X_n$ and $\tilde{X}_n$ are defined on the same probability space
$\Omega_n$. If $X_n\geq\tilde{X}_n$ and $X_n\Rightarrow D$
as well as $\tilde{X}_n\Rightarrow D$, then $X_n-\tilde{X}_n$ converges
to zero in probability.
Conversely if $\tilde{X}_n\Rightarrow D$ and $X_n-\tilde{X}_n$
converges to zero in probability then $X_n\Rightarrow D$ as well.
\end{lem}

Finally we come to the proof of Theorem~\ref{ThmAsymptSpaceTime}.

\begin{pf*}{Proof of Theorem~\ref{ThmAsymptSpaceTime}}
Let us define the random variables
%
\begin{equation}
\Xi_k:=X_t^{\mathrm{resc}}(u_k,
\theta_k)-X_t^{\mathrm{resc}}(u_k,0)
\end{equation}
such that
%
\begin{equation}
\label{eq5.14} \qquad \Pb \Biggl(\bigcap_{k=1}^m
\bigl\{X_t^{\mathrm{resc}}(u_k,\theta_k)\leq
s_k\bigr\} \Biggr) = \Pb \Biggl(\bigcap_{k=1}^m
\bigl\{X_t^{\mathrm{resc}}(u_k,0)+\Xi_k\leq
s_k\bigr\} \Biggr).
\end{equation}
The slow-decorrelation result (Proposition~\ref{PropSlowDecorrelation})
implies $\Xi_k\to0$ in probability as $t\to\infty$. Introducing $\e>0$
we can use inclusion--exclusion to decompose
%
\begin{equation}\qquad
 \eqref{eq5.14}=\Pb \Biggl(\bigcap
_{k=1}^m\bigl\{X_t^{\mathrm{resc}}(u_k,0)+
\Xi _k\leq s_k\bigr\}\cap\bigl\{|\Xi_k|\leq\e\bigr\}
\Biggr)+\sum_j \Pb (R_j ),\hspace*{-10pt}
\end{equation}
where the sum on the right-hand side is finite, and each $R_j$
satisfies $R_j\subset\{|\Xi_k|> \e\}$ for at least one $k$.
Using the limit result from Theorem~\ref{thmAsympFixedTime},
%
\begin{equation}
\lim_{t\to\infty}\Pb \Biggl(\bigcap_{k=1}^m
\bigl\{X_t^{\mathrm{resc}}(u_k,0)\leq s_k
\bigr\} \Biggr)=\Pb \Biggl(\bigcap_{k=1}^m
\bigl\{{\cal A}_1(u_k)\leq s_k\bigr\}
\Biggr),
\end{equation}
leads to
%
\begin{eqnarray}
\limsup_{t\to\infty}\Pb \Biggl(\bigcap
_{k=1}^m\bigl\{X_t^{\mathrm{resc}}(u_k,
\theta _k)\leq s_k\bigr\} \Biggr) &\leq&\Pb \Biggl(\bigcap
_{k=1}^m\bigl\{{\cal A}_1(u_k)
\leq s_k+\e\bigr\} \Biggr),
\nonumber
\\[-8pt]
\\[-8pt]
\nonumber
\liminf_{t\to\infty}\Pb \Biggl(\bigcap_{k=1}^m
\bigl\{X_t^{\mathrm{resc}}(u_k,\theta _k)\leq
s_k\bigr\} \Biggr) &\geq&\Pb \Biggl(\bigcap
_{k=1}^m\bigl\{{\cal A}_1(u_k)
\leq s_k-\e\bigr\} \Biggr).
\end{eqnarray}
Since the joint distribution function of the Airy$_1$ process is
continuous in $s_1,\ldots,s_m$, we can take the limit $\e\to0$ and obtain
%
\begin{equation}
\lim_{t\to\infty}\Pb \Biggl(\bigcap_{k=1}^m
\bigl\{X_t^{\mathrm{resc}}(u_k,\theta _k)\leq
s_k\bigr\} \Biggr) = \Pb \Biggl(\bigcap_{k=1}^m
\bigl\{{\cal A}_1(u_k)\leq s_k\bigr\}
\Biggr).
\end{equation}
\upqed\end{pf*}

\begin{appendix}\label{app}
\section{Bounds on the Lambert W function}\label{AppendixLambert}

\begin{lem}[(Path of $\Gamma^\rho$ and its image under $\varphi
$)]\label
{lemContour}
For any $\rho\in[0,1)$ the contour $\Gamma^\rho=\{\gamma(\tau
)=L_{\lfloor\tau\rfloor}(-e^{2\pi\I\tau-1}(1-\rho)),\tau\in\R
\setminus
[0,1)\}$, with $L_k(z)$ being the $k$th branch of the Lambert W
function, satisfies:
\begin{longlist}[(10)]
\item[(1)] $\Gamma^\rho$ crosses the real line at one unique
$z_0\leq-1$.
\item[(2)] $z_0=-1-\sqrt{2\rho}+\Or(\rho)$.
\item[(3)] $\Re(z)<\Re(z_0)$ for all $z\in\Gamma^\rho\setminus\{
z_0\}$.
\item[(4)] $\Re(z)$ is monotone along each part of $\Gamma^\rho
\setminus
\{z_0\}$.
\item[(5)] $\llvert \frac{\mathrm{d}}{\mathrm{d}\tau}\Re(\gamma(\tau))\rrvert \leq
3\pi$ for
$|\tau|\geq2$.
\item[(6)] $\Gamma^\rho$ has asymptotic angle $\pm\pi/2$.

In addition,
\item[(7)] $\varphi(z)$ crosses the real line infinitely often at the
two unique points \mbox{$z_0^*=\varphi(z_0)\geq-1$} and $z_1^*>z_0^*$
when $z$ moves along $\Gamma^\rho$.
\item[(8)] $z_0^*=-1$ if and only if $\rho=0$.
\item[(9)] $\Re(\varphi(z))>\Re(\varphi(z_0))$ for all $z\in\Gamma
^\rho$
with $\varphi(z)\neq\varphi(z_0)$.
\item[(10)] $\Re(z)$ is monotone along each part of $\varphi
(\Gamma
^\rho )\setminus\{z_0^*,z_1^*\}$.
\end{longlist}
\end{lem}

\begin{lem}[(Behavior of $f_3$ along $\Gamma^\rho$)]\label{lemf3}
The function $f_3(z)=(z+1)^2-(\varphi(z)+1)^2$ satisfies:
\begin{longlist}[(1)]
\item[(1)] $f_3 (\Gamma^\rho )$ crosses the real line at one
unique $\widehat{z}_0=f_3(z_0)$, where $z_0$ is given as in Lemma~\ref
{lemContour}.
\item[(2)] $\widehat{z}_0=0$ if $\rho=0$.
\item[(3)] $\Re(f_3(z))<\Re(f_3(z_0))$ for all $z\in\Gamma^\rho
\setminus
\{z_0\}$.
\item[(4)] $\Re(f_3(z))$ is monotone along each part of $f_3
(\Gamma
^\rho )\setminus\{\widehat{z}_0\}$.
\item[(5)] $\llvert \frac{\mathrm{d}}{\mathrm{d}\tau}\Re(f_3(\gamma(\tau)))\rrvert \geq4\pi
^2|\tau|$ for $|\tau|\geq5$.
\end{longlist}
\end{lem}

\begin{pf*}{Proof of Lemma~\ref{lemContour}}
Write $\gamma(\tau)=L_{\lfloor\tau\rfloor} (-e^{2\pi\I\tau
-1}(1-\rho
) )$. The branch cut of the Lambert function is done in such a
way that
%
\begin{eqnarray}
\label{eqImPart} %
(2k-2)\pi&\leq&\Im \bigl(L_k(z)
\bigr)\leq(2k+1)\pi\qquad\mbox{for } k>0,
\nonumber\\
-\pi&\leq&\Im \bigl(L_k(z) \bigr)\leq\pi\qquad\mbox{for } k=0,
\\
(2k-1)\pi&\leq&\Im \bigl(L_k(z) \bigr)\leq(2k+2)\pi\qquad\mbox{for }
k<0; \nonumber
\end{eqnarray}
see also Figure~4 of~\cite{CGHJK96}. The curve $\gamma(\tau)$ changes
branches every time when $\tau\in\Z$, but since at each jump point the
function $-e^{2\pi\I\tau-1}(1-\rho)$ meets the line $(-\infty,0]$,
which is the location of the branch cut, the function $\gamma(\tau)$ is
in fact continuous at these points, and $\Gamma^\rho$ therefore connected.

The function $L_k(z)$ satisfies the differential identity ((3.2)
in~\cite{CGHJK96})
%
\begin{equation}
L_k'(z)=\frac{L_k(z)}{z(1+L_k(z))}.
\end{equation}
By elementary calculus we therefore have
%
\begin{equation}
\label{eqDiffId} \gamma'(\tau)=\frac{\mathrm{d}}{\mathrm{d}\tau}\gamma(\tau)=2\pi\I
\biggl(1-\frac
{1}{\gamma
(\tau)+1} \biggr).
\end{equation}
From the structure of the branches one has \mbox{$\lim_{\tau\searrow
1}\gamma(\tau)=\lim_{\tau\nearrow0}\gamma(\tau)\leq-1$}. This
limit is
our $z_0$. The image of $-1/e$ under $L_{-1}$, $L_0$ and $L_1$ is $-1$,
so $z_0=-1$, which corresponds to $\rho=0$.

Consider first $\tau>1$. Since all the involved branches lie in the
upper half plane, we have $\Im(\gamma(\tau))>0$. Additionally, the fact
that the transformations $z\mapsto z+1$ and $z\mapsto-z^{-1}$ map the
upper half plane onto itself implies by \eqref{eqDiffId} the inequality
$\Re(\gamma'(\tau))<0$. This in turn implies \mbox{$\Re(\gamma
(\tau
))\leq-1$} which can be inserted in \eqref{eqDiffId}, leading to $\Im
(\gamma'(\tau))\geq2\pi$. So for $\tau>1$ the curve $\gamma(\tau
)$ is
moving monotone north-west in $\tau$. Analogously we can argue that
$\gamma(\tau)$ is moving monotone south-west in $|\tau|$ for $\tau<0$.

Thereby the claims (1), (3) and (4) are settled. To see claim~(6), we notice
that for large $|\tau|$ also $|\gamma(\tau)|$ is large and the fraction
in \eqref{eqDiffId} tends to zero, resulting in $\gamma'(\tau)\to
2\pi\I$.

By \eqref{eqImPart} we have $|\Im(\gamma(\tau))|\geq2\pi$ for all
$|\tau
|\geq2$. Inserting this in \eqref{eqDiffId} results in $|\gamma
'(\tau
)|\leq2\pi(1+1/2\pi)\leq3\pi$ and consequently claim~(5).

Again by~\cite{CGHJK96} the series \eqref{LambTayl} is the expansion of
$L_1$ or $L_{-1}$, respectively, when inserting $p(z)=-\sqrt{2(ez+1)}$
instead of $p(z)=\sqrt{2(ez+1)}$. Which branch one gets depends on the
sign of $\Im(z)$. Claim~(2) follows.

For the corresponding statements on $\varphi(z)$, first notice the identity
%
\begin{equation}
\label{eqPhiFrac} \varphi\bigl(\gamma(\tau)\bigr)=L_0 \bigl(\gamma(
\tau)e^{\gamma(\tau
)} \bigr)=L_0 \bigl(-e^{2\pi\I\tau-1}(1-\rho)
\bigr)=\gamma\bigl(\tau-\lfloor \tau \rfloor\bigr),
\end{equation}
from which it is clear that $\varphi(\gamma(\tau))$ is periodic in
$\tau
$. We can therefore reduce our considerations to $\tau\in[0,1)$.

By (4.4) of~\cite{CGHJK96}, the principal branch of the Lambert W
function is given by
%
\begin{equation}
\bigl\{a+\I b\in\C, a+b\cot(b)>0 \mbox{ and } -\pi<b<\pi\bigr\}.
\end{equation}
So regarding points of the principal branch, by
%
\begin{equation}
\sgn\Im \bigl((a+\I b)e^{a+\I b} \bigr)=\sgn (a\sin b +b\cos b )=\sgn b,
\end{equation}
the function $z\mapsto ze^z$ preserves the sign of the imaginary part.
But then its inverse function $L_0$ must do the same. Consequently,
$\Im
(\gamma(\tau))<0$ for $0<\tau<1/2$ and $\Im(\gamma(\tau))>0$ for
$1/2<\tau<1$. In the same way as before this leads through \eqref
{eqDiffId} to $\Re(\gamma'(\tau))>0$ for $0<\tau<1/2$ and $\Re
(\gamma
'(\tau))<0$ for $1/2<\tau<1$. This settles claims (7), (9) and (10) with
$z_1^*=\gamma(1/2)$.

The equation $z_0^*=-1$ is equivalent to $L_0 (-e^{-1}(1-\rho
)
)=-1$, which clearly holds for $\rho=0$, and by injectivity in the
principal branch for no other $\rho$.
\end{pf*}

\begin{pf*}{Proof of Lemma~\ref{lemf3}}
With $\{\tau\}=\tau-\lfloor\tau\rfloor$ being the fractional part of
$\tau$ we write using \eqref{eqPhiFrac}
%
\begin{equation}
f_3\bigl(\gamma(\tau)\bigr)=\bigl(\gamma(\tau)+1
\bigr)^2-\bigl(\gamma\bigl(\{\tau\}\bigr)+1\bigr)^2.
\end{equation}
Differentiating with respect to $\tau$ results in
%
\begin{equation}
\label{df3} \frac{\mathrm{d}}{\mathrm{d}\tau}f_3\bigl(\gamma(\tau)\bigr)=4\pi\I
\bigl(\gamma(\tau )-\gamma\bigl(\{\tau \}\bigr) \bigr).
\end{equation}
By Lemma~\ref{lemContour} we know that $\Re (\gamma(\tau
)-\gamma(\{
\tau\}) )<0$ which gives\break  $\Im\frac{\mathrm{d}}{\mathrm{d}\tau}f_3(\gamma(\tau))<0$.
The monotonicity of the imaginary part entails the uniqueness in claim~(1).

Regarding the real part, first notice that for $\tau\nearrow0$ or
$\tau
\searrow1$, $\Im (\gamma(\tau)-\gamma(\{\tau\})
)$ tends
to zero, resulting in $\Re\frac{\mathrm{d}}{\mathrm{d}\tau}f_3(\gamma(\tau))=0$. By
differentiating a second time we arrive at
%
\begin{equation}
\frac{\mathrm{d}^2}{\mathrm{d}\tau^2}f_3\bigl(\gamma(\tau)\bigr)=8\pi^2
\biggl(\frac{1}{\gamma
(\tau
)+1}-\frac{1}{\gamma(\{\tau\})+1} \biggr).
\end{equation}
From Lemma~\ref{lemContour} the right-hand side has negative real part.
Integrating results in the desired monotonicity and therefore claims~(3)
and (4).

By \eqref{eqImPart} we have $|\Im(\gamma(\tau))|\geq2\pi(|\tau
|-2)$ for
all $\tau\in\R$. Combining this with $|\Im(\gamma(\{\tau\}))|\leq
\pi$
results in \mbox{$|\Im(\gamma(\tau)-\gamma(\{\tau\}))|\geq\pi
|\tau|$}
for $|\tau|\geq5$. With \eqref{df3}, claim~(5) follows.

Claim~(2) is a corollary of Lemma~\ref{lemContour}, claim~(2).
\end{pf*}

\section{Correlation kernel for determinantal measures}\label{AppendixA}
Here are two useful results from~\cite{BFPS06}. They are written for
the continuous case. The proofs are identical to the discrete case.

\begin{lem}[(See Lemma~3.3 of~\cite{BFPS06})]\label{AppLemma3.3} Let $f$
an antisymmetric function of $\{x_1^{N},\ldots,x_N^{N}\}$. Then,
whenever $f$ has enough decay to make the sums finite,
%
\begin{equation}
\int_{\cal D} f\bigl(x_1^{N},
\ldots,x_N^{N}\bigr) \prod_{2\leq i\leq j \leq N}
\,\D x_i^j=\int_{\cal D'} f
\bigl(x_1^{N},\ldots,x_N^{N}\bigr)
\prod_{2\leq i\leq j
\leq N} \,\D x_i^j,
\end{equation}
where
%
\begin{eqnarray}
{\cal D}&=&\bigl\{x_i^j,2 \leq i \leq j \leq N |
x_i^j>x_i^{j+1},x_i^j
\geq x_{i-1}^{j-1}\bigr\},
\nonumber
\\[-8pt]
\\[-8pt]
\nonumber
{\cal D'}&=&\bigl\{x_i^j,2 \leq i \leq
j \leq N | x_i^j\geq x_{i-1}^{j-1}
\bigr\},
\end{eqnarray}
and the positions $x_1^1>x_1^2>\cdots>x_1^N$ being fixed.
\end{lem}

\begin{lem}[(See Lemma~3.4 of~\cite{BFPS06})]\label{AppLemma2}
Assume we have a signed measure on $\{x_i^n,n=1,\ldots,N,i=1,\ldots,n\}
$ given in the form
%
\begin{equation}
\label{Sasweight} \frac{1}{Z_N}\prod_{n=1}^{N-1}
\det\bigl[\phi_n\bigl(x_i^n,x_j^{n+1}
\bigr)\bigr]_{1\leq
i,j\leq n+1} \det\bigl[\Psi_{N-i}^{N}
\bigl(x_{j}^N\bigr)\bigr]_{1\leq i,j \leq N},
\end{equation}
where $x_{n+1}^n$ are some ``virtual'' variables and $Z_N$ is a
normalization constant. If $Z_N\neq0$, then the correlation functions
are determinantal.

To write down the kernel we need to introduce some notation. Define
%
\begin{equation}
\label{Sasdef phi12} \phi^{(n_1,n_2)}(x,y)= %
\cases{ (\phi_{n_1}
\ast\cdots\ast\phi_{n_2-1}) (x,y),&\quad $n_1<n_2$,
\vspace*{2pt}
\cr
0,&\quad $n_1\geq n_2$,} %
\end{equation}
where $(a* b)(x,y)=\int_\R\,\D z\, a(x,z) b(z,y)$, and, for $1\leq n<N$,
%
\begin{equation}
\label{Sasdef_psi} \Psi_{n-j}^{n}(x):= \bigl(
\phi^{(n,N)} * \Psi_{N-j}^{N}\bigr) (y), \qquad j=1,\ldots,N.
\end{equation}
Set $\phi_0(x_1^0,x)=1$. Then the functions
%
\begin{equation}\qquad
\bigl\{ \bigl(\phi_0*\phi^{(1,n)}\bigr)
\bigl(x_1^0,x\bigr), \ldots,\bigl(\phi_{n-2}*
\phi ^{(n-1,n)}\bigr) \bigl(x_{n-1}^{n-2},x\bigr),
\phi_{n-1}\bigl(x_{n}^{n-1},x\bigr)\bigr\}
\end{equation}
are linearly independent and generate the $n$-dimensional space $V_n$.
Define a set of functions $\{\Phi_j^{n}(x), j=0,\ldots,n-1\}$ spanning
$V_n$ defined by the orthogonality relations
%
\begin{equation}
\label{Sasortho} \int_\R\,\D x \Phi_i^n(x)
\Psi_j^n(x) = \delta_{i,j}
\end{equation}
for $0\leq i,j\leq n-1$.

Further, if $\phi_n(x_{n+1}^n,x)=c_n \Phi_0^{(n+1)}(x)$, for some
$c_n\neq0$, \mbox{$n=1,\ldots,N-1$}, then the kernel takes the
simple form
%
\begin{equation}
\label{SasK} \qquad K(n_1,x_1;n_2,x_2)=
-\phi^{(n_1,n_2)}(x_1,x_2)+ \sum
_{k=1}^{n_2} \Psi _{n_1-k}^{n_1}(x_1)
\Phi_{n_2-k}^{n_2}(x_2).
\end{equation}
\end{lem}
\end{appendix}

\section*{Acknowledgments}
The authors thank Neil O'Connell for discussions on reflected Brownian motions
and Ivan Corwin for pointing at the extension to general initial
conditions. T. Weiss is grateful to Neil O'Connell and Nikos Zygouras
for their hospitality at Warwick University.

%



\printaddresses

\begin{thebibliography}{36}

\bibitem{AO76}
%
\begin{barticle}[author]
\bauthor{\bsnm{Anderson},~\bfnm{R.~F.}\binits{R.~F.}} \AND
\bauthor{\bsnm{Orey},~\bfnm{S.}\binits{S.}}
(\byear{1976}).
\btitle{Small random perturbation of dynamical systems with
reflecting boundary}.
\bjournal{Nagoya Math. J.}
\bvolume{60}
\bpages{189--216}.
\end{barticle}
%
\bptok{imsref}%
\endbibitem

\bibitem{BJ00}
%
\begin{barticle}[author]
\bauthor{\bsnm{Banwell},~\bfnm{T.~C.}\binits{T.~C.}} \AND
\bauthor{\bsnm{Jayakumar},~\bfnm{A.}\binits{A.}}
(\byear{2000}).
\btitle{Exact analytical solution for current flow through
diode with series resistance}.
\bjournal{Electronics Letters}
\bvolume{36}
\bpages{291--292}.
\end{barticle}
%
\bptok{imsref}%
\endbibitem

\bibitem{BPLPCS00}
%
\begin{barticle}[mr]
\bauthor{\bsnm{Barry},~\bfnm{D.~A.}\binits{D.~A.}},
\bauthor{\bsnm{Parlange},~\bfnm{J.-Y.}\binits{J.-Y.}},
\bauthor{\bsnm{Li},~\bfnm{L.}\binits{L.}},
\bauthor{\bsnm{Prommer},~\bfnm{H.}\binits{H.}},
\bauthor{\bsnm{Cunningham},~\bfnm{C.~J.}\binits{C.~J.}} \AND
\bauthor{\bsnm{Stagnitti},~\bfnm{F.}\binits{F.}}
(\byear{2000}).
\btitle{Analytical approximations for real values of the {L}ambert
{$W$}-function}.
\bjournal{Math. Comput. Simulation}
\bvolume{53}
\bpages{95--103}.
\bid{doi={10.1016/S0378-4754(00)00172-5}, issn={0378-4754}, mr={1777736}}
\end{barticle}
%
\bptok{imsref}%
\endbibitem

\bibitem{Bar01}
%
\begin{barticle}[mr]
\bauthor{\bsnm{Baryshnikov},~\bfnm{Yu.}\binits{Yu.}}
(\byear{2001}).
\btitle{G{UE}s and queues}.
\bjournal{Probab. Theory Related Fields}
\bvolume{119}
\bpages{256--274}.
\bid{doi={10.1007/PL00008760}, issn={0178-8051}, mr={1818248}}
\end{barticle}
%
\bptok{imsref}%
\endbibitem

\bibitem{BC09}
%
\begin{barticle}[mr]
\bauthor{\bsnm{Ben Arous},~\bfnm{G{\'e}rard}\binits{G.}} \AND
\bauthor{\bsnm{Corwin},~\bfnm{Ivan}\binits{I.}}
(\byear{2011}).
\btitle{Current fluctuations for {TASEP}: A proof of the {P}r\"
ahofer--{S}pohn conjecture}.
\bjournal{Ann. Probab.}
\bvolume{39}
\bpages{104--138}.
\bid{doi={10.1214/10-AOP550}, issn={0091-1798}, mr={2778798}}
\end{barticle}
%
\bptok{imsref}%
\endbibitem

\bibitem{BC11}
%
\begin{barticle}[author]
\bauthor{\bsnm{Borodin},~\bfnm{A.}\binits{A.}} \AND
\bauthor{\bsnm{Corwin},~\bfnm{I.}\binits{I.}}
(\byear{2014}).
\btitle{Macdonald processes.}
\bjournal{Probab. Theory Related Fields}
\bvolume{158}
\bpages{225--400}.
\end{barticle}
%
\bptok{imsref}%
\endbibitem

\bibitem{BF07}
%
\begin{barticle}[mr]
\bauthor{\bsnm{Borodin},~\bfnm{Alexei}\binits{A.}} \AND
\bauthor{\bsnm{Ferrari},~\bfnm{Patrik~L.}\binits{P.~L.}}
(\byear{2008}).
\btitle{Large time asymptotics of growth models on space-like paths.
{I}. {P}ush{ASEP}}.
\bjournal{Electron. J. Probab.}
\bvolume{13}
\bpages{1380--1418}.
\bid{doi={10.1214/EJP.v13-541}, issn={1083-6489}, mr={2438811}}
\end{barticle}
%
\bptok{imsref}%
\endbibitem

\bibitem{BFP06}
%
\begin{barticle}[author]
\bauthor{\bsnm{Borodin},~\bfnm{A.}\binits{A.}},
\bauthor{\bsnm{Ferrari},~\bfnm{P.~L.}\binits{P.~L.}} \AND
\bauthor{\bsnm{Pr{\"a}hofer},~\bfnm{M.}\binits{M.}}
(\byear{2007}).
\btitle{{Fluctuations in the discrete TASEP with periodic initial
configurations and the Airy$_1$ process}}.
\bjournal{Int. Math. Res. Papers}
\bvolume{2007}
\bpages{rpm002}.
\end{barticle}
%
\bptok{imsref}%
\endbibitem

\bibitem{BFPS06}
%
\begin{barticle}[mr]
\bauthor{\bsnm{Borodin},~\bfnm{Alexei}\binits{A.}},
\bauthor{\bsnm{Ferrari},~\bfnm{Patrik~L.}\binits{P.~L.}},
\bauthor{\bsnm{Pr{\"a}hofer},~\bfnm{Michael}\binits{M.}} \AND
\bauthor{\bsnm{Sasamoto},~\bfnm{Tomohiro}\binits{T.}}
(\byear{2007}).
\btitle{Fluctuation properties of the {TASEP} with periodic initial
configuration}.
\bjournal{J. Stat. Phys.}
\bvolume{129}
\bpages{1055--1080}.
\bid{doi={10.1007/s10955-007-9383-0}, issn={0022-4715}, mr={2363389}}
\end{barticle}
%
\bptok{imsref}%
\endbibitem

\bibitem{BFS07}
%
\begin{barticle}[mr]
\bauthor{\bsnm{Borodin},~\bfnm{Alexei}\binits{A.}},
\bauthor{\bsnm{Ferrari},~\bfnm{Patrik~L.}\binits{P.~L.}} \AND
\bauthor{\bsnm{Sasamoto},~\bfnm{Tomohiro}\binits{T.}}
(\byear{2008}).
\btitle{Transition between {${\mathrm{Airy}}_1$} and {${\mathrm{Airy}}_2$}
processes and {TASEP} fluctuations}.
\bjournal{Comm. Pure Appl. Math.}
\bvolume{61}
\bpages{1603--1629}.
\bid{doi={10.1002/cpa.20234}, issn={0010-3640}, mr={2444377}}
\end{barticle}
%
\bptok{imsref}%
\endbibitem

\bibitem{CY92}
%
\begin{barticle}[mr]
\bauthor{\bsnm{Chang},~\bfnm{Chih~Chung}\binits{C.~C.}} \AND
\bauthor{\bsnm{Yau},~\bfnm{Horng-Tzer}\binits{H.-T.}}
(\byear{1992}).
\btitle{Fluctuations of one-dimensional {G}inzburg--{L}andau models in
nonequilibrium}.
\bjournal{Comm. Math. Phys.}
\bvolume{145}
\bpages{209--234}.
\bid{issn={0010-3616}, mr={1162798}}
\end{barticle}
%
\bptok{imsref}%
\endbibitem

\bibitem{Che02}
%
\begin{barticle}[mr]
\bauthor{\bsnm{Chen},~\bfnm{YangQuan}\binits{Y.}} \AND
\bauthor{\bsnm{Moore},~\bfnm{Kevin~L.}\binits{K.~L.}}
(\byear{2002}).
\btitle{Analytical stability bound for delayed second-order systems
with repeating poles using {L}ambert function {$W$}}.
\bjournal{Automatica J. IFAC}
\bvolume{38}
\bpages{891--895}.
\bid{doi={10.1016/S0005-1098(01)00264-3}, issn={0005-1098}, mr={2133607}}
\end{barticle}
%
\bptok{imsref}%
\endbibitem

\bibitem{CGHJ93}
%
\begin{barticle}[author]
\bauthor{\bsnm{Corless},~\bfnm{R.~M.}\binits{R.~M.}},
\bauthor{\bsnm{Gonnet},~\bfnm{G.~H.}\binits{G.~H.}},
\bauthor{\bsnm{Hare},~\bfnm{D.~E.~G.}\binits{D.~E.~G.}} \AND
\bauthor{\bsnm{Jeffrey},~\bfnm{D.~J.}\binits{D.~J.}}
(\byear{1993}).
\btitle{Lambert's {W} {f}unction in {M}aple}.
\bjournal{The Maple Technical Newsletter}
\bvolume{9}
\bpages{12--22}.
\end{barticle}
%
\bptok{imsref}%
\endbibitem

\bibitem{CGHJK96}
%
\begin{barticle}[author]
\bauthor{\bsnm{Corless},~\bfnm{R.~M.}\binits{R.~M.}},
\bauthor{\bsnm{Gonnet},~\bfnm{G.~H.}\binits{G.~H.}},
\bauthor{\bsnm{Hare},~\bfnm{D.~E.~G.}\binits{D.~E.~G.}},
\bauthor{\bsnm{Jeffrey},~\bfnm{D.~J.}\binits{D.~J.}} \AND
\bauthor{\bsnm{Knuth},~\bfnm{D.~E.}\binits{D.~E.}}
(\byear{1996}).
\btitle{{On the Lambert W function}}.
\bjournal{Adv. Comput. Math.}
\bvolume{5}
\bpages{329--359}.
\end{barticle}
%
\bptok{imsref}%
\endbibitem

\bibitem{CJK97}
%
\begin{binproceedings}[mr]
\bauthor{\bsnm{Corless},~\bfnm{Robert~M.}\binits{R.~M.}},
\bauthor{\bsnm{Jeffrey},~\bfnm{David~J.}\binits{D.~J.}} \AND
\bauthor{\bsnm{Knuth},~\bfnm{Donald~E.}\binits{D.~E.}}
(\byear{1997}).
\btitle{A sequence of series for the {L}ambert {$W$} function}.
In \bbooktitle{Proceedings of the 1997 {I}nternational {S}ymposium on
{S}ymbolic and {A}lgebraic {C}omputation ({K}ihei, {HI})}
\bpages{197--204 (electronic)}.
\bpublisher{ACM},
\blocation{New York}.
\bid{doi={10.1145/258726.258783}, mr={1809988}}
\end{binproceedings}
%
\bptok{imsref}%
\endbibitem

\bibitem{CJV00}
%
\begin{barticle}[author]
\bauthor{\bsnm{Corless},~\bfnm{R.~M.}\binits{R.~M.}},
\bauthor{\bsnm{Jeffrey},~\bfnm{D.~J.}\binits{D.~J.}} \AND
\bauthor{\bsnm{Valluri},~\bfnm{S.~R.}\binits{S.~R.}}
(\byear{2000}).
\btitle{Some applications of the {L}ambert {W} function to physics}.
\bjournal{Canadian Journal of Physics}
\bvolume{78}
\bpages{823--831}.
\end{barticle}
%
\bptok{imsref}%
\endbibitem

\bibitem{CFP10b}
%
\begin{barticle}[mr]
\bauthor{\bsnm{Corwin},~\bfnm{Ivan}\binits{I.}},
\bauthor{\bsnm{Ferrari},~\bfnm{Patrik~L.}\binits{P.~L.}} \AND
\bauthor{\bsnm{P{\'e}ch{\'e}},~\bfnm{Sandrine}\binits{S.}}
(\byear{2012}).
\btitle{Universality of slow decorrelation in {KPZ} growth}.
\bjournal{Ann. Inst. Henri Poincar\'e Probab. Stat.}
\bvolume{48}
\bpages{134--150}.
\bid{doi={10.1214/11-AIHP440}, issn={0246-0203}, mr={2919201}}
\end{barticle}
%
\bptok{imsref}%
\endbibitem

\bibitem{Fer08}
%
\begin{barticle}[author]
\bauthor{\bsnm{Ferrari},~\bfnm{P.~L.}\binits{P.~L.}}
(\byear{2008}).
\btitle{{Slow decorrelations in KPZ growth}}.
\bjournal{J. Stat. Mech.}
\bvolume{2008}
\bpages{P07022}.
\end{barticle}
%
\bptok{imsref}%
\endbibitem

\bibitem{FN08b}
%
\begin{barticle}[author]
\bauthor{\bsnm{Forrester},~\bfnm{P.~J.}\binits{P.~J.}} \AND
\bauthor{\bsnm{Nagao},~\bfnm{T.}\binits{T.}}
(\byear{2011}).
\btitle{{Determinantal correlations for classical projection processes}}.
\bjournal{J. Stat. Mech.}
\bvolume{2011}
\bpages{P08011}.
\end{barticle}
%
\bptok{imsref}%
\endbibitem

\bibitem{Har65}
%
\begin{barticle}[mr]
\bauthor{\bsnm{Harris},~\bfnm{T.~E.}\binits{T.~E.}}
(\byear{1965}).
\btitle{Diffusion with ``collisions'' between particles}.
\bjournal{J. Appl. Probab.}
\bvolume{2}
\bpages{323--338}.
\bid{issn={0021-9002}, mr={0184277}}
\end{barticle}
%
\bptok{imsref}%
\endbibitem

\bibitem{HW87}
%
\begin{barticle}[mr]
\bauthor{\bsnm{Harrison},~\bfnm{J.~M.}\binits{J.~M.}} \AND
\bauthor{\bsnm{Williams},~\bfnm{R.~J.}\binits{R.~J.}}
(\byear{1987}).
\btitle{Multidimensional reflected {B}rownian motions having
exponential stationary distributions}.
\bjournal{Ann. Probab.}
\bvolume{15}
\bpages{115--137}.
\bid{issn={0091-1798}, mr={0877593}}
\end{barticle}
%
\bptok{imsref}%
\endbibitem

\bibitem{JK04}
%
\begin{barticle}[author]
\bauthor{\bsnm{Jain},~\bfnm{A.}\binits{A.}} \AND
\bauthor{\bsnm{Kapoor},~\bfnm{A.}\binits{A.}}
(\byear{2004}).
\btitle{Exact analytical solutions of the parameters of real solar
cells using {L}ambert {W}-function}.
\bjournal{Solar Energy Materials and Solar Cells}
\bvolume{81}
\bpages{269--277}.
\end{barticle}
%
\bptok{imsref}%
\endbibitem

\bibitem{KPS12}
%
\begin{bmisc}[author]
\bauthor{\bsnm{Karatzas},~\bfnm{I.}\binits{I.}},
\bauthor{\bsnm{Pal},~\bfnm{S.}\binits{S.}} \AND
\bauthor{\bsnm{Shkolnikov},~\bfnm{M.}\binits{M.}}
(\byear{2012}).
\bhowpublished{Systems of Brownian particles with asymmetric collisions.
Available at \arxivurl{arXiv:1210.0259}.}
\end{bmisc}
%
\bptok{imsref}%
\endbibitem

\bibitem{KPZ86}
%
\begin{barticle}[author]
\bauthor{\bsnm{Kardar},~\bfnm{M.}\binits{M.}},
\bauthor{\bsnm{Parisi},~\bfnm{G.}\binits{G.}} \AND
\bauthor{\bsnm{Zhang},~\bfnm{Y.~Z.}\binits{Y.~Z.}}
(\byear{1986}).
\btitle{Dynamic scaling of growing interfaces}.
\bjournal{Phys. Rev. Lett.}
\bvolume{56}
\bpages{889--892}.
\end{barticle}
%
\bptok{imsref}%
\endbibitem

\bibitem{ledDIL}
%
\begin{bincollection}[mr]
\bauthor{\bsnm{Ledoux},~\bfnm{M.}\binits{M.}}
(\byear{2007}).
\btitle{Deviation inequalities on largest eigenvalues}.
In \bbooktitle{Geometric Aspects of Functional Analysis}.
\bseries{Lecture Notes in Math.}
\bvolume{1910}
\bpages{167--219}.
\bpublisher{Springer},
\blocation{Berlin}.
\bid{doi={10.1007/978-3-540-72053-9_10}, mr={2349607}}
\end{bincollection}
%
\bptok{imsref}%
\endbibitem

\bibitem{OCon09}
%
\begin{barticle}[mr]
\bauthor{\bsnm{O'Connell},~\bfnm{Neil}\binits{N.}}
(\byear{2012}).
\btitle{Directed polymers and the quantum {T}oda lattice}.
\bjournal{Ann. Probab.}
\bvolume{40}
\bpages{437--458}.
\bid{doi={10.1214/10-AOP632}, issn={0091-1798}, mr={2952082}}
\end{barticle}
%
\bptok{imsref}%
\endbibitem

\bibitem{OCY01}
%
\begin{barticle}[mr]
\bauthor{\bsnm{O'Connell},~\bfnm{Neil}\binits{N.}} \AND
\bauthor{\bsnm{Yor},~\bfnm{Marc}\binits{M.}}
(\byear{2001}).
\btitle{Brownian analogues of {B}urke's theorem}.
\bjournal{Stochastic Process. Appl.}
\bvolume{96}
\bpages{285--304}.
\bid{doi={10.1016/S0304-4149(01)00119-3}, issn={0304-4149}, mr={1865759}}
\end{barticle}
%
\bptok{imsref}%
\endbibitem

\bibitem{Sas05}
%
\begin{barticle}[author]
\bauthor{\bsnm{Sasamoto},~\bfnm{T.}\binits{T.}}
(\byear{2005}).
\btitle{Spatial correlations of the {1D KPZ} surface on a flat substrate}.
\bjournal{J. Phys. A}
\bvolume{38}
\bpages{L549--L556}.
\end{barticle}
%
\bptok{imsref}%
\endbibitem

\bibitem{SW98}
%
\begin{barticle}[mr]
\bauthor{\bsnm{Sasamoto},~\bfnm{Tomohiro}\binits{T.}} \AND
\bauthor{\bsnm{Wadati},~\bfnm{Miki}\binits{M.}}
(\byear{1998}).
\btitle{Determinant form solution for the derivative nonlinear {S}chr\"
odinger type model}.
\bjournal{J. Phys. Soc. Japan}
\bvolume{67}
\bpages{784--790}.
\bid{doi={10.1143/JPSJ.67.784}, issn={0031-9015}, mr={1621629}}
\end{barticle}
%
\bptok{imsref}%
\endbibitem

\bibitem{Sch97}
%
\begin{barticle}[mr]
\bauthor{\bsnm{Sch{\"u}tz},~\bfnm{Gunter~M.}\binits{G.~M.}}
(\byear{1997}).
\btitle{Exact solution of the master equation for the asymmetric
exclusion process}.
\bjournal{J. Stat. Phys.}
\bvolume{88}
\bpages{427--445}.
\bid{doi={10.1007/BF02508478}, issn={0022-4715}, mr={1468391}}
\end{barticle}
%
\bptok{imsref}%
\endbibitem

\bibitem{Sko61}
%
\begin{barticle}[author]
\bauthor{\bsnm{Skorokhod},~\bfnm{A.~V.}\binits{A.~V.}}
(\byear{1961}).
\btitle{Stochastic equations for diffusions in a bounded region}.
\bjournal{Theory Probab. Appl.}
\bvolume{6}
\bpages{264--274}.
\end{barticle}
%
\bptok{imsref}%
\endbibitem

\bibitem{TW94}
%
\begin{barticle}[mr]
\bauthor{\bsnm{Tracy},~\bfnm{Craig~A.}\binits{C.~A.}} \AND
\bauthor{\bsnm{Widom},~\bfnm{Harold}\binits{H.}}
(\byear{1994}).
\btitle{Level-spacing distributions and the {A}iry kernel}.
\bjournal{Comm. Math. Phys.}
\bvolume{159}
\bpages{151--174}.
\bid{issn={0010-3616}, mr={1257246}}
\end{barticle}
%
\bptok{imsref}%
\endbibitem

\bibitem{TW96}
%
\begin{barticle}[mr]
\bauthor{\bsnm{Tracy},~\bfnm{Craig~A.}\binits{C.~A.}} \AND
\bauthor{\bsnm{Widom},~\bfnm{Harold}\binits{H.}}
(\byear{1996}).
\btitle{On orthogonal and symplectic matrix ensembles}.
\bjournal{Comm. Math. Phys.}
\bvolume{177}
\bpages{727--754}.
\bid{issn={0010-3616}, mr={1385083}}
\end{barticle}
%
\bptok{imsref}%
\endbibitem

\bibitem{WW84}
%
\begin{barticle}[mr]
\bauthor{\bsnm{Varadhan},~\bfnm{S.~R.~S.}\binits{S.~R.~S.}} \AND
\bauthor{\bsnm{Williams},~\bfnm{R.~J.}\binits{R.~J.}}
(\byear{1985}).
\btitle{Brownian motion in a wedge with oblique reflection}.
\bjournal{Comm. Pure Appl. Math.}
\bvolume{38}
\bpages{405--443}.
\bid{doi={10.1002/cpa.3160380405}, issn={0010-3640}, mr={0792398}}
\bptnote{check year}%
\end{barticle}
%
\bptok{imsref}%
\endbibitem

\bibitem{War07}
%
\begin{barticle}[mr]
\bauthor{\bsnm{Warren},~\bfnm{Jon}\binits{J.}}
(\byear{2007}).
\btitle{Dyson's {B}rownian motions, intertwining and interlacing}.
\bjournal{Electron. J. Probab.}
\bvolume{12}
\bpages{573--590}.
\bid{doi={10.1214/EJP.v12-406}, issn={1083-6489}, mr={2299928}}
\end{barticle}\vadjust{\eject}
%
\bptok{imsref}%
\endbibitem

\bibitem{Wei11}
%
\begin{bmisc}[author]
\bauthor{\bsnm{Weiss},~\bfnm{T.}\binits{T.}}
(\byear{2011}).
\bhowpublished{Scaling behaviour of the directed polymer model of
Baryshnikov and O'Connell at zero temperature.
Bachelor thesis, TU-M{\"u}nchen}.
\end{bmisc}
%
\bptok{imsref}%
\endbibitem

\end{thebibliography}
\end{document}